\documentclass[aps,prd,superscriptaddress,twocolumn,nofootinbib]{revtex4-2}%
\usepackage{soul}
\usepackage[colorlinks=true,
            linkcolor=blue,
            citecolor=blue,
            urlcolor=blue]{hyperref}

\usepackage{booktabs}   

\usepackage[english]{babel}
\usepackage[utf8x]{inputenc}
\usepackage[T1]{fontenc}
\usepackage{orcidlink}

\usepackage{amsmath}
\usepackage{amssymb}
\usepackage{graphicx}
\usepackage[usenames,dvipsnames]{xcolor}
\usepackage{url}
\usepackage{mmacells}

\usepackage{caption}      
\usepackage{color}        

\usepackage{tikz}
\usetikzlibrary{positioning}

\usepackage[normalem]{ulem} 

\usepackage{chngcntr} 

\newcommand{\dd}{\textrm{d}}

\begin{document}

\title{Fisher and DALI posterior approximations for testing gravity with gravitational waves}

\author{Felipe A. da Silva Barbosa}
\affiliation{PPGCosmo, Universidade Federal do Espírito Santo, 29075-910 Vitória–ES, Brazil}
\affiliation{Cosmo-Ufes, Universidade Federal do Espírito Santo, 29075-910 Vitória–ES, Brazil}

\author{Davi C. Rodrigues\orcidlink{0000-0003-1683-5443}}
\affiliation{PPGCosmo, Universidade Federal do Espírito Santo, 29075-910 Vitória–ES, Brazil}
\affiliation{Cosmo-Ufes, Universidade Federal do Espírito Santo, 29075-910 Vitória–ES, Brazil}
\affiliation{Departamento de Física, Universidade Federal do Espírito Santo, 29075-910 Vitória–ES, Brazil}

\author{Josiel Mendonça Soares de Souza\orcidlink{0000-0003-1552-0095}}
\affiliation{PPGCosmo, Universidade Federal do Espírito Santo, 29075-910 Vitória–ES, Brazil}
\affiliation{Cosmo-Ufes, Universidade Federal do Espírito Santo, 29075-910 Vitória–ES, Brazil}

\author{Miguel Quartin\orcidlink{0000-0001-5853-6164}}
\affiliation{PPGCosmo, Universidade Federal do Espírito Santo, 29075-910 Vitória–ES, Brazil}
\affiliation{Centro Brasileiro de Pesquisas Físicas, 22290-180, Rio de Janeiro, RJ, Brazil}
\affiliation{Observatório do Valongo, Universidade Federal do Rio de Janeiro, 20080-090, Rio de Janeiro, RJ, Brazil}

\date{\today }

\begin{abstract}
We investigate the accuracy and computational efficiency of posterior approximations for gravitational-wave (GW) parameter estimation based on Fisher Matrix and its higher-order extension, the Derivative Approximation for LIkelihoods (DALI), in the context of tests of General Relativity using parameterized deviations in the inspiral waveforms (the TIGER formalism). We adopt the \texttt{IMRPhenomD} waveform and compare the exact posterior sampling to the following approximations in increasing order of complexity: the traditional Fisher Matrix, the Fisher sampling (singlet-DALI), the doublet-DALI and the triplet-DALI. We find that both singlet and doublet-DALI are significantly faster than the exact posterior by factors between 20 and 1000, depending on the inspiral SNRs -- larger gains correspond to larger SNRs. These gains are considerably larger for more complete waveforms, such as \texttt{IMRPhenomHM}. To evaluate the accuracy of the approximations, we use Jensen-Shannon divergence. We conclude that both singlet and doublet-DALI systematically improves over the standard Fisher Matrix approximation, while triplet-DALI shows large variability, limiting its practical advantage. We also introduce a code with a new implementation of Fisher and DALI analysis (Symbolic DALI, \texttt{SymDALI}) which computes derivatives symbolically via \textit{Wolfram Language}. 
\end{abstract}

\maketitle

\section{Introduction}

The Einstein Telescope (ET) \cite{Punturo:2010zz, ET:2025xjr} and Cosmic Explorer (CE) \cite{Evans:2021gyd} will expand our gravitational wave (GW) data dramatically, observing mergers at unprecedented rates and signal-to-noise ratios (SNRs). For instance, the number of expected binary black hole (BBH) coalescences per year reaches tens of thousands, with over a thousand with SNR exceeding 100~\cite{Branchesi:2023mws}.  This large dataset will allow for very precise tests of fundamental physics, cosmology, the population of both black holes and neutron stars, and more (see~\cite{ET:2025xjr} for a review of the science case of ET). On the other hand, even in standard General Relativity the computational challenge of performing a full Bayesian analysis for each event in these catalogs is large, as each waveform is described by more than ten parameters. Allowing for different gravity and/or cosmological theories only increases the computational complexity. This motivates the investigation of approximate Bayesian methods for GW analysis.

In physics in general, one of the most commonly used approximation tool in the past three decades is the Fisher matrix (FM) formalism~\cite{Tegmark:1997rp,Sellentin:2014zta}.
It is particular commonly used in GW forecasts~\cite{Vallisneri:2007ev}, with different alternative codes available in the literature~\cite{Borhanian:2020ypi,Dupletsa:2022scg,Iacovelli:2022mbg,Begnoni:2025oyd, deSouza:2025qok}. The FM assumes a posterior which is Gaussian in all parameters. It is exact if the likelihood is Gaussian in the data and the model is linear on the parameters. If the model is non-linear, it should still provide a reasonable estimate if the uncertainties are small, since it relies on an expansion of the posterior around the best-fit. In any case, the approximation is expected to improve as the SNR increases.

The FM formalism introduces major computational advantages as all information needed for parameter estimation is contained in a single matrix, the inversion of which determines the confidence intervals and correlations of all parameters. It also allows for trivial marginalization of any number of parameters, as well as fixing a posteriori any number of parameters. However, there are two main difficulties involved. The first is that for parameter sets involving ${\cal O}(10)$ or more parameters, such as those in GW analysis, numerical instability may be a concern due to the presence of almost degenerate degrees of freedom. This problem can be avoided with the use of auto-differentiation methods~\cite{autodiff1964} and/or with addition of mildly informed priors.
The second, which is harder to solve, is that the most reasonable priors are often non-Gaussian, which cannot be implemented in the standard FM approach (e.g., \cite{Rodriguez:2013mla}). This can be a problem for GW analysis even in high SNR cases.

The Derivative Approximation for LIkelihoods (DALI)~\cite{Sellentin:2014zta} extends the Fisher matrix approach\footnote{We employ the expected FM (also called frequentist FM~\cite{Sellentin:2014zta}). This is the approach commonly used in forecasting, where the FM is defined as an expectation value, $F_{ij} = \langle \partial_i \mathcal{L} \, \partial_j \mathcal{L} \rangle$, where $\mathcal{L}$ is the log-likelihood, $\partial_i$ denotes the derivative with respect to the $i$-th model parameter, and the expectation is taken over the data distribution $p(d|\theta_{\rm fid})$ corresponding to the fiducial model~\cite{EfronHinkley1978}.} by including higher-order terms in a Taylor expansion of the log-likelihood around its maximum. A direct Taylor expansion is ill-defined, and leads to non-normalizable probabilities. However, it was shown in~\cite{Sellentin:2014zta} that by assuming that the observational data follow a Gaussian distribution, 
a reordering of the Taylor expansion in terms of derivative powers of the theory vector, as opposed to grouping at powers of the perturbation parameter, ensures that the resulting series defines a valid probability function (see below for more details). This assumption of data Gaussianity is consistent with the standard framework for the analysis of gravitational wave data with short signals~\cite{Finn:1992wt, Cutler:1994ys, Vallisneri:2007ev, Wang:2022kia}. This may not be the case for all signals detected by third generation detectors, since longer signals are more likely to suffer from transient noise  \cite{Emma:2026urt}.
The first term beyond the Gaussian FM approximation is called the doublet-DALI, followed by the triplet-DALI and so on. The main drawback of using DALI, when compared to the standard FM, is that marginalization over parameters can no longer be performed by simply dropping rows and columns of a matrix. Instead, the full approximate posterior must still be sampled in the whole multi-dimensional parameter space, and numerically integrated over all parameters one wish to marginalize over. It also requires the calculation of a larger number of derivatives.

In the context of GW, the DALI method was previously investigated in~\cite{Wang:2022kia} and~\cite{desouza:2023ozp}. 
Recently, it was revisited  in~\cite{deSouza:2025qok}, where a detailed study of the accuracy and computational cost advantages of both doublet-DALI and triplet-DALI was performed, and the public \texttt{GWDALI} code version 1.0 was released.\footnote{\url{https://github.com/jmsdsouzaphd/gwdali}}
One of the important findings in~\cite{deSouza:2025qok} is that performing a numerical sampling of the posterior has major advantages over the FM counterpart because it can naturally include any kind of analytical non-Gaussian priors, as already discussed in~\cite{Dupletsa:2022scg, desouza:2023ozp}. See also \cite{Dupletsa:2024gfl} and references therein for validations against real data from the Gravitational Wave Transient Catalogs. This dramatically improves the accuracy of the approximation for GW analyses even for the simplest DALI approximation (the singlet-DALI, which can also be called sampled Fisher), which is otherwise equivalent to the Fisher expansion (see also \cite{Tagliazucchi:2026dpr}).
Despite these advantages, DALI has not yet been applied to tests of GR and fundamental physics using GWs. 
In particular, a well-established framework for testing gravity and fundamental physics is through the parametrization of the waveform with the phenomenological inclusion of additional post-Newtonian parameters. A proposed framework is the so-called Test Infrastructure for GEneral Relativity (TIGER)~\cite{Agathos:2013upa, Roy:2025gzv},  in which fractional deviations $\delta\varphi_i$ to the inspiral phase coefficients are introduced --- more details below. This approach has been tested in recent LVK collaboration studies~\cite{LIGOScientific:2025obp, LIGOScientific:2026fcf}, and has the advantage of being both simple and agnostic with respect to particular modified gravity models.

In this work, we investigate the use of DALI in the TIGER formalism. Using the BBH catalog created in~\cite{Branchesi:2023mws} and the IMRPhenomD waveform model~\cite{Husa:2015iqa, Khan:2015jqa}, we compare five inference methods: exact likelihood, FM, singlet-DALI and doublet-DALI. We also comment on our findings for triplet-DALI.
Following~\cite{deSouza:2025qok} we quantify the accuracy using the Jensen-Shannon divergence (JSD) between the exact one-dimensional posteriors and each different approximation methods. We also introduce a new code implementation of Fisher and DALI: Symbolic DALI (\texttt{SymDALI}) \cite{SymDALI}. \texttt{SymDALI} is capable of performing Fisher and DALI analysis for the usual parameters of GR, besides being capable of also dealing with the \texttt{TIGER} formalism for the \texttt{IMRPhenomD} waveform. 
We discuss the code in more detail in Appendix~\ref{symdali}. We point out that a comparison between FM results and GW data in the context of \texttt{TIGER} was done in \cite{Begnoni:2025mtz}, which also performed a forecast for ET considering different detector geometries.

\section{Model setup}

\subsection{DALI}
The DALI method provides a positive and normalizable approximation for the likelihood, and constitutes a higher-order extension of the Fisher matrix approximation. It assumes Gaussian-distributed observational data, and it has a simpler form when the covariance is parameter-independent. This is precisely the case for GW parameter inference, where the likelihood is given by
\cite{Finn:1992wt, Cutler:1994ys, Vallisneri:2007ev} (see also \cite{Wang:2022kia}),
\begin{equation}
  p(s|\theta) \propto \exp\left[-\frac{1}{2} \left\langle s-h(\theta) \mid s-h(\theta) \right \rangle  \right] .
\end{equation}
In the above, $s$ stands for the GW signal and $h(\theta)$ for the template waveform, which depends nonlinearly on the parameters $\theta$. The inner product between two real time-domain functions $h(t)$ and $g(t)$ is defined as
\begin{equation}
  \langle h|g \rangle = 2 \int_0^\infty  \frac{\tilde h^*(f)\tilde g(f) + \tilde h(f)\tilde g^*(f)}{S_n(f)}
  \, \dd f .
\end{equation}
Here $S_n(f)$ is the one-sided noise power spectral density, and the tilde denotes a Fourier transform.

The GW likelihoods are described by $N$ parameters $\boldsymbol{\theta} = \{\theta_1, \dots, \theta_N\}$. In the DALI approach we write the log likelihood as the following expansion around the best fit (BF) (see~\cite{Wang:2022kia, deSouza:2025qok} for more details):
{\footnotesize
\begin{align}\label{Eq:DALI}    
    & \log\mathcal{L}  \;=\;
    \log\mathcal{L}_{\rm BF} - \left[\frac{1}{2}\sum_{i,j}\langle \partial_i h| \partial_j h \rangle \Delta \theta^{ij}\right]_{\rm BF} \nonumber\\
    &  { - \left[\frac{1}{2}\sum_{i,j,k}\langle \partial_i h | \partial_j\partial_k h \rangle\Delta\theta^{ijk}
         +\frac{1}{8}\sum_{i,j,k,l}\langle \partial_i\partial_j h | \partial_k\partial_l h \rangle\Delta\theta^{ijkl}\right]_{\rm BF}} \nonumber \\
    & {- \left[\frac{1}{6}\!\!\sum_{i,j,k,l}\!\!\langle \partial_i h | \partial_j\partial_k\partial_l h \rangle\Delta\theta^{ijkl}
         \!+\! \frac{1}{12}\!\!\sum_{i,...,m}\!\!\langle \partial_i\partial_j h | \partial_k\partial_l\partial_m h \rangle\Delta\theta^{ijklm}\right.}\nonumber \\
    &\quad\;\; \left. + \,\frac{1}{72} \sum_{i,...,n}\langle \partial_i\partial_j\partial_k h | \partial_l\partial_m\partial_n h \rangle\Delta\theta^{ijklmn} \right]_{\rm BF}\,,
\end{align}}%
where $\Delta\theta^i\equiv \theta^i-\theta^i_{BF}$ and $\Delta\theta^{i\dots k} \equiv \Delta\theta^i \dots \Delta\theta^k$. The first bracket corresponds to the Fisher matrix approximation (singlet-DALI), the second to the doublet-DALI contribution and the last one to the extra terms from the triplet-DALI. 
It is theoretically simple to go beyond the triplet-DALI approximation written in Eq~\eqref{Eq:DALI}. However each additional order increases the computation time, here we focus on lower order terms and reserve higher order investigations for future work. We follow \cite{deSouza:2025qok} and calculate derivatives in $y=1/d_L$, where $d_L$ is the luminosity distance.

\subsection{TIGER formalism}

Compact-object mergers provide a unique laboratory for testing the validity of GR in a strong-field and highly dynamical regime. In fact, gravitational-wave observations are used to perform a variety of consistency tests of GR. These range from carrying out parameter estimation independently on the inspiral and ringdown portions of the waveform and checking for consistency in the inferred remnant properties, to so-called null tests, in which the underlying waveform model is modified such that agreement with GR corresponds to a set of additional parameters being equal to zero \cite{LIGOScientific:2026fcf}. This work focuses on null-tests in the context of the \texttt{TIGER} framework

Given the inspiral phase,
\begin{align}\label{inspiral_phase}
    &\phi_{\text{TF2}}(f)  =  2 \pi f t_c - \phi_c -\pi/4   \nonumber \\
    & \quad + \frac{3}{128 \eta} v^{-5} \sum^7_{i=0}\left[ \varphi_i + \varphi_i^{(l)} \ln(v^3) \right]v^i,
\end{align}
where $v\equiv (\pi GMf)^{1/3}$, the parameters $\varphi_i$ are general functions of the binary's intrinsic parameters 
\begin{align}\label{Extra Pn}
    \varphi_i \equiv \varphi_i^{NS} (\eta)  + \varphi_i( \eta, \chi_1, \chi_2),
\end{align}
where the upper index $\rm NS$ denotes the non-spinning part of the Post-Newtonian (PN) coefficients.
We follow the \texttt{TIGER} formalism and the tests performed by the LVK collaboration \cite{LIGOScientific:2019fpa, LIGOScientific:2020tif} (see also \cite{ET:2025xjr}), by adding extra parameters in the non-spinning contribution
\begin{align}
    \varphi_i^{NS} (\eta) \rightarrow (1+ \delta \varphi_i)  \varphi^{NS}_i (\eta).
\end{align}
The baseline approximant used here is \texttt{IMRPhenomD}. We observe that, in this notation, PN counting is equivalent to $i/2$: $\varphi_1$ corresponds to 0.5PN and so on. In Appendix \ref{app_comp} we compare the implementation of \texttt{IMRPhenomD} in \texttt{SymDALI} and on \texttt{LAL} \cite{Wette:2020air}.

\subsection{Detector Network}

We assume the 15 km,  2L-misaligned configuration for ET and one L shaped CE of 40 km. The ET detectors lie on the same plane but are rotated with respect to each other.  We place the ET detectors in Sardinia and the Meuse Rhine region -- see section 2 in \cite{Branchesi:2023mws}. The CE detector is placed on New Mexico -- see Table III in \cite{Borhanian:2020ypi}. For the ASD's we assume ET working with cryogenic technology and the baseline sensitivity for CE. We take all detectors to have an 85\% independent duty cycle, meaning that in 15\% of the time the detectors are assumed to be off for maintenance.

We assume a sampling frequency of 2048 Hz, implying that the maximum frequency used in Fourier space is 1024 Hz, and a minimum frequency $f_{\text{min}}=5$ Hz for all 3 detectors. The reference frequency is taken to be $f_{\text{min}}$, and the frequency bins in Fourier space are fixed by the duration of the signal $\Delta f= 1/\Delta T$.



\subsection{Population}

\begin{table}
\caption{Selection criteria applied to the BBH gravitational-wave signals here analyzed. We observe that subindex `net' denotes the network and IMR the inspiral-merger-ringdown}
\centering
\begin{tabular}{p{4cm} p{4cm}}
\hline
\hline
Criterion & Requirement \\
\hline
IMR SNR$_{\text{net}}$ & $\mathrm{SNR} > 100$ \\
Inspiral SNR$_{\text{net}}$ & $\mathrm{SNR}_{\mathrm{inspiral}} > 12$ \\
Per-detector SNR & $\mathrm{SNR}_{\mathrm{det}} > 8$ \\
\hline
\hline
\end{tabular}
\label{tab:selection_criteria}
\end{table}

\begin{table*}
    \caption{
    Adopted prior distributions. The superscript ``fid'' stands for fiducial.  The interval limits for $t_c$ and $d_{\rm L}$ are technical choices to speed up convergence (i.e., larger limits yield the same results). The cosmology is Planck 2018~\cite{Planck:2018vyg}. As usual, the parameters $\theta$, $\phi$, $t_c$ and $\psi$ are in Earth reference frame.
    }
    \centering
    \renewcommand{\arraystretch}{1.4} 
    \begin{tabular}{ c c} 
    \hline
    \hline
    Parameter &  Prior\\ 
    \hline
    $\theta$ (polar angle) & $\cos(\theta) \sim U(-1,1)$ \\ 
    $\phi$ (azimuthal angle) &  $\phi \sim U(0, 2\pi)$ \\
    $\theta_{\text{JN}}$ (inclination) & $\cos(\iota)$ $\sim U(-1, 1)$ \\
    $\phi_{\text{c}}$  (coalescence phase) & $\phi_{\text{c}} \sim U(0, 2\pi)$\\
    $\psi$ (polarization angle) & $\psi \sim U(0, \pi)$\\
    $t_c$ (coalescence time) & $t_c-t_c^{\rm fid} \sim U(-0.1 {\rm s}, 0.1 {\rm s})$ \\
    $\chi_i$ (aligned spins) &  $p(\chi_i) = -\frac{1}{2 \times 0.99} \ln(|\chi_i|), \chi_i\in (-0.99, 0.99)$\\
    $d_L$ (lumin. dist.) & $\frac{1}{1+z} \frac{\dd V_c}{\dd z}$ , [$d^{\text{fid}}_L-10 \mbox{ Gpc}$, $d^{\text{fid}}_L+10 \mbox{ Gpc}$] \\
    $(\mathcal{M}_c, \, q)$ (chirp mass and mass ratio) & $\mathcal{M}_c  \, \frac{(1+q)^{2/5}}{q^{6/5}}$ see Appendix \ref{app_prior}\\
    \hline
    \hline
    \end{tabular} 
    \label{distr external params}
\end{table*}

Our baseline BBH population catalog is the one developed in \cite{Mapelli:2021gyv}.\footnote{The data is available at \url{https://apps.et-gw.eu/tds/?content=3&r=18321}.} This catalog generates the intrinsic properties of a BBH system by evolving binary stars until they become binary BBH's and it also takes into account dynamical formation channels, that is the formation of BBH's by encounters of black holes in globular clusters and hierarchical formation channels. The final catalog contains $1.2 \times 10^5$ BBH mergers over a period of one year. 

For testing GR in the inspiral phase, we focus on events that satisfy the conditions on Table \ref{tab:selection_criteria}.\footnote{We define the cutoff frequency of the inspiral to be $f_{\text{cut}} = 0.018/(G M)$, which follows from the frequency cutoff for the inspiral phase ansatz in \texttt{IMRPhenomD} \cite{Khan:2015jqa}.} We observe that there are around $10^4$ events that satisfy these conditions. 

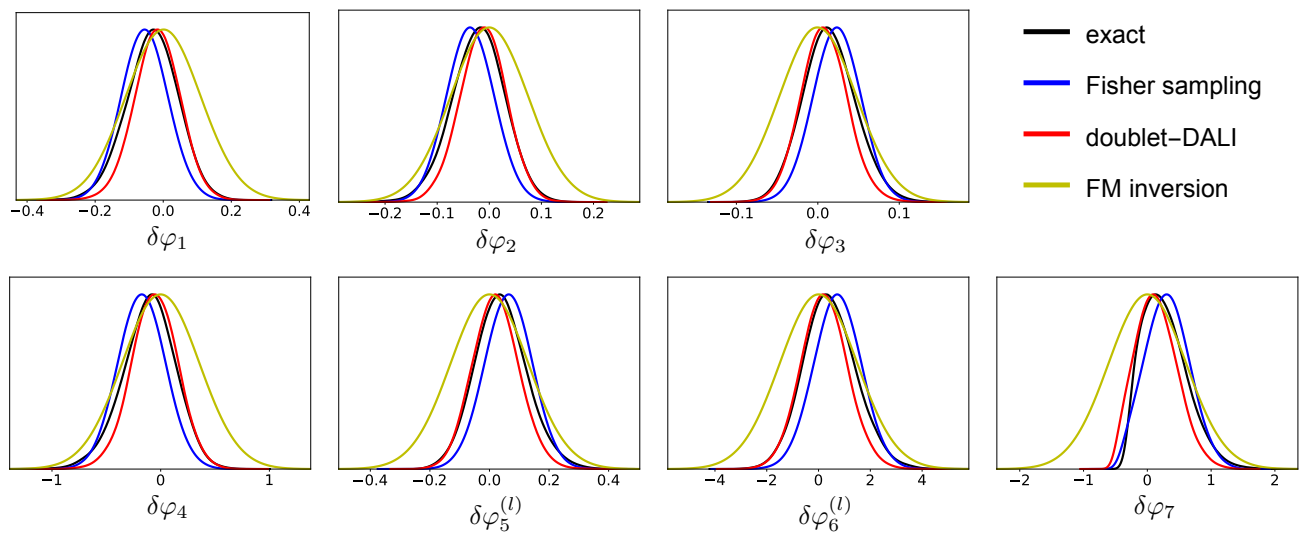
\begin{figure*}
\begin{tikzpicture}

\node(img0) {\includegraphics[width=0.23\textwidth]{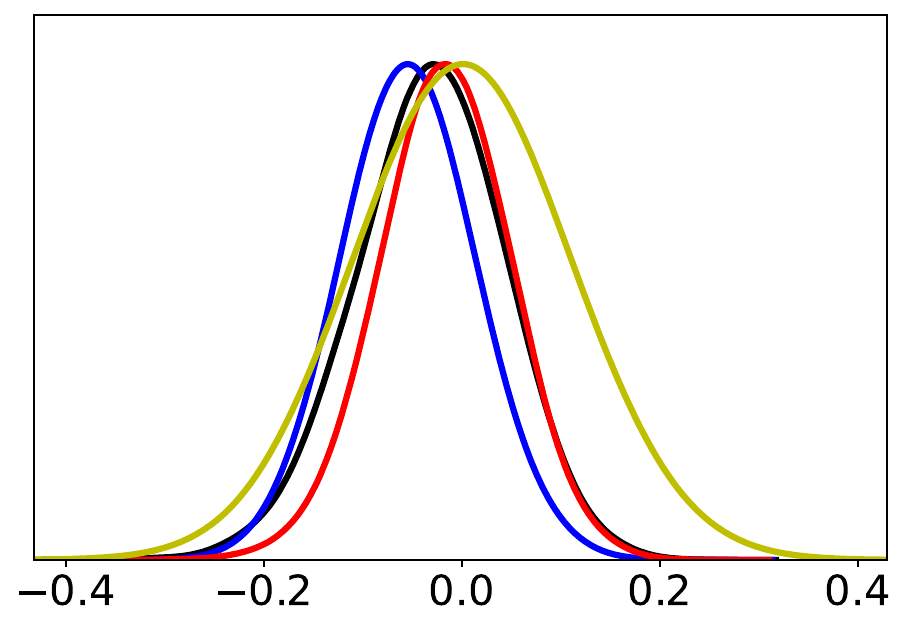}};
\node[right=of img0, xshift=-1cm] (img1) {\includegraphics[width=0.23\textwidth]{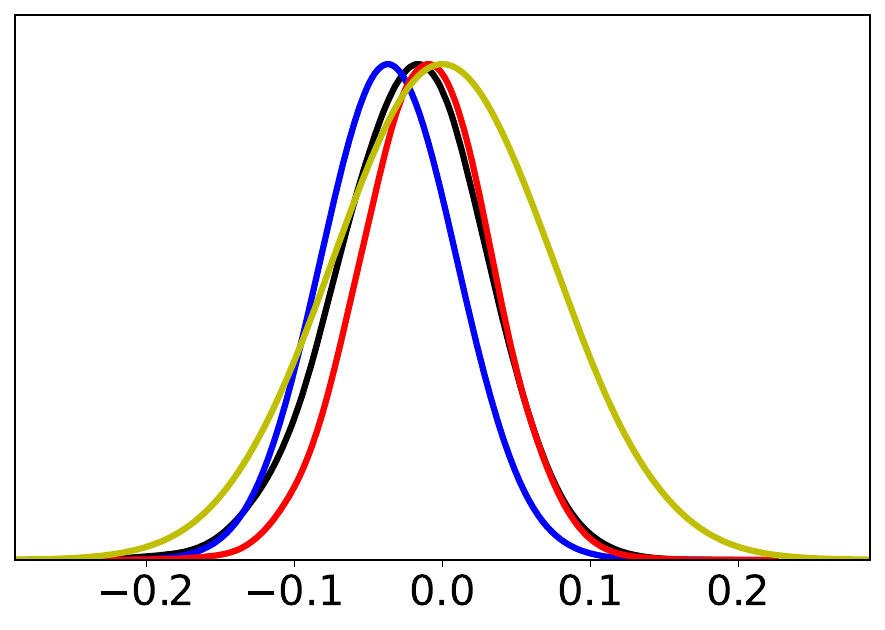}};
\node[right=of img1, xshift=-1cm] (img2) {\includegraphics[width=0.23\textwidth]{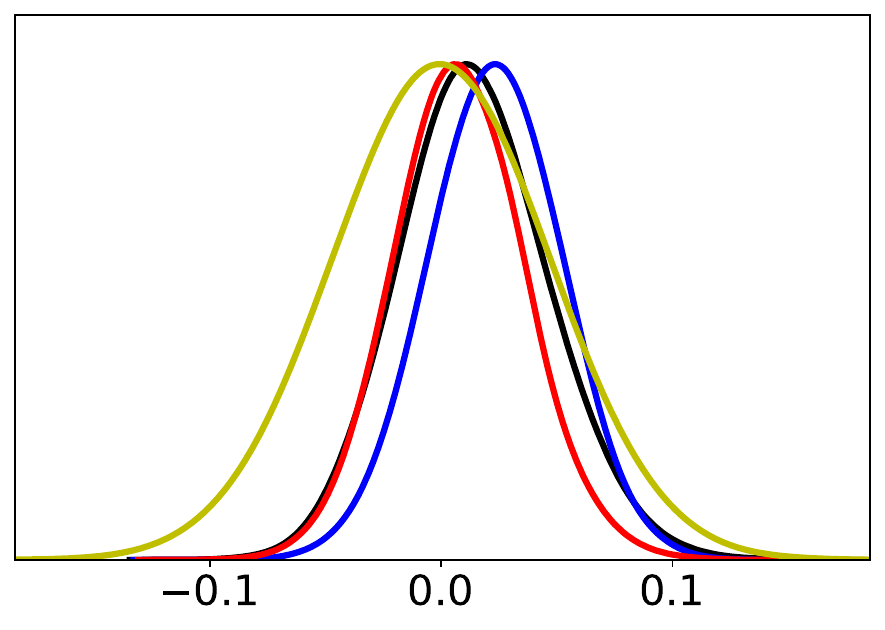}};
\node[right=of img2, xshift=-0.95cm, yshift=0.05cm] (img2b) {\includegraphics[width=0.24\textwidth]{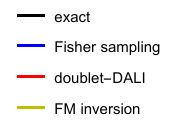}};

\node[below=of img0, node distance=0cm, yshift=1.2cm, xshift=0.1cm,
      font=\fontsize{10}{10}\selectfont] {$\delta \varphi_1$};
\node[below=of img1, node distance=0cm, yshift=1.2cm, xshift=0.1cm,
      font=\fontsize{10}{10}\selectfont] {$\delta \varphi_2$};
\node[below=of img2, node distance=0cm, yshift=1.2cm, xshift=0.1cm,
      font=\fontsize{10}{10}\selectfont] {$\delta \varphi_3$};

\node[below=of img0, yshift=0.6cm] (img3) {\includegraphics[width=0.23\textwidth]{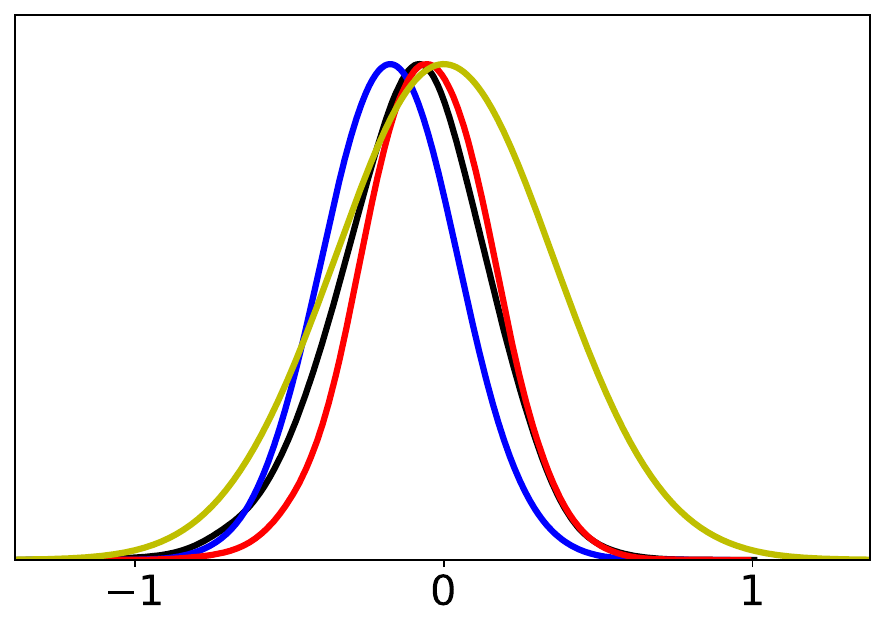}};
\node[right=of img3, xshift=-1cm] (img4) {\includegraphics[width=0.23\textwidth]{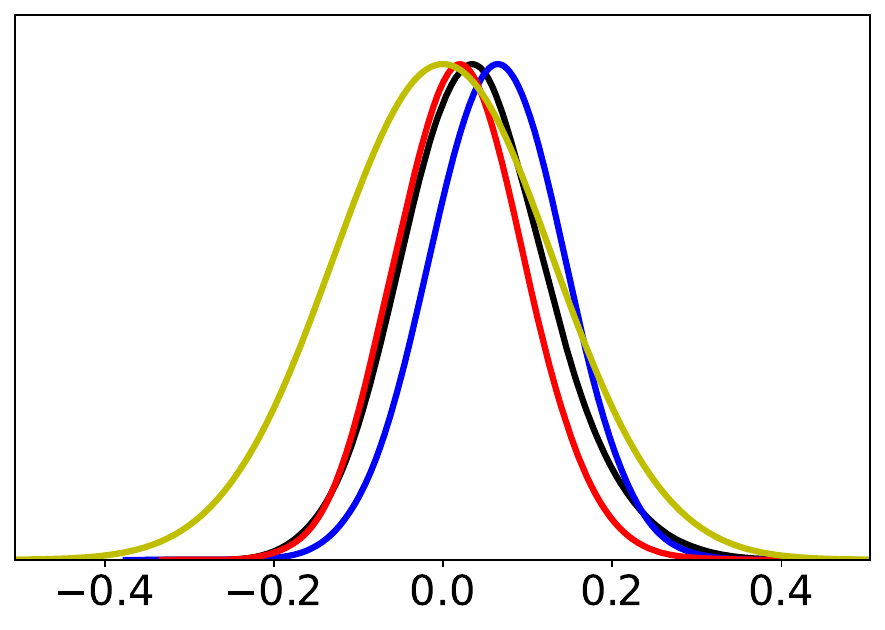}};
\node[right=of img4, xshift=-1cm] (img5) {\includegraphics[width=0.23\textwidth]{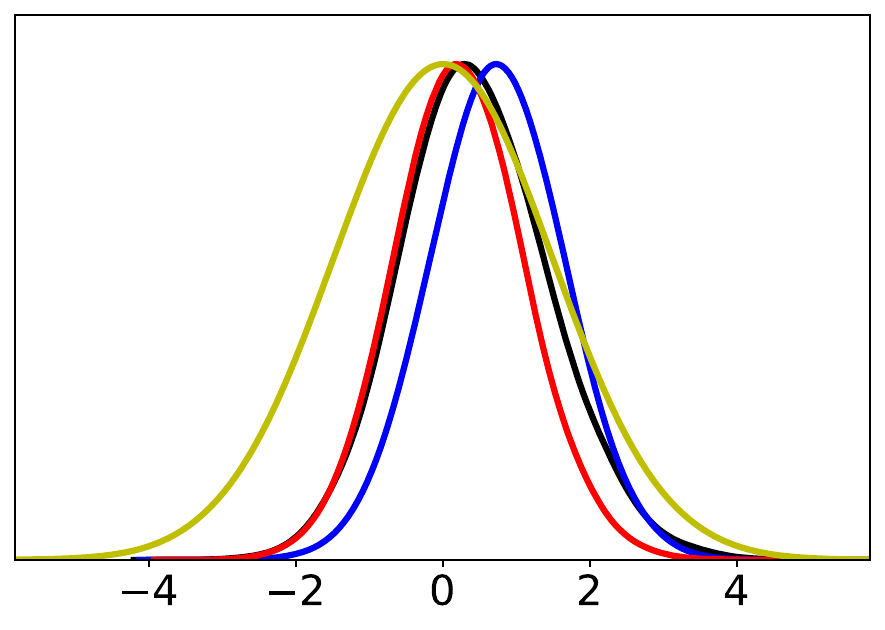}};
\node[right=of img5, xshift=-1cm] (img6) {\includegraphics[width=0.23\textwidth]{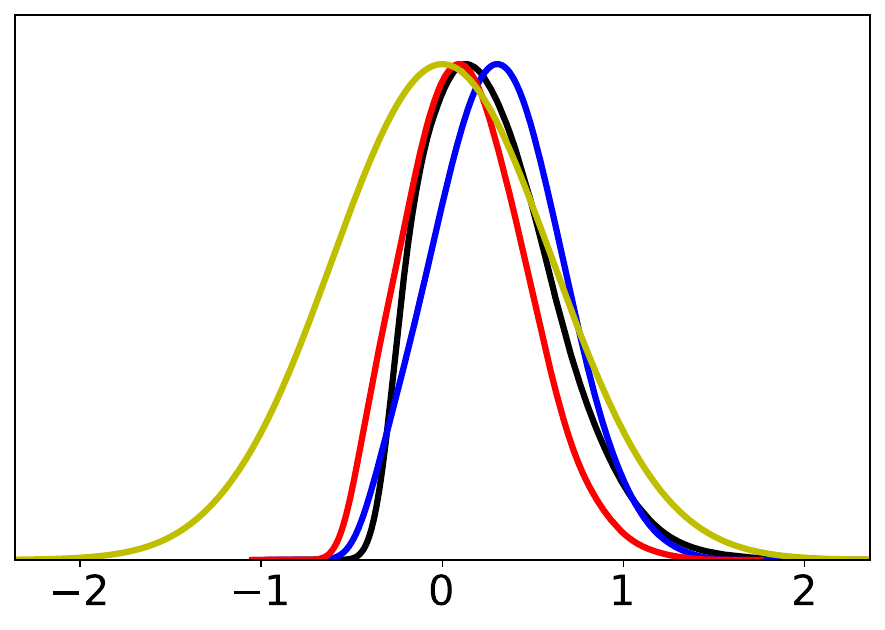}};

\node[below=of img3, node distance=0cm, yshift=1.2cm, xshift=0.1cm,
      font=\fontsize{10}{10}\selectfont] {$\delta \varphi_4$};
\node[below=of img4, node distance=0cm, yshift=1.2cm, xshift=0.1cm,
      font=\fontsize{10}{10}\selectfont] {$\delta \varphi^{(l)}_5$};
\node[below=of img5, node distance=0cm, yshift=1.2cm, xshift=0.1cm,
      font=\fontsize{10}{10}\selectfont] {$\delta \varphi^{(l)}_6$};
\node[below=of img6, node distance=0cm, yshift=1.2cm, xshift=0.1cm,
      font=\fontsize{10}{10}\selectfont] {$\delta \varphi_7$};


\end{tikzpicture}
\caption{Representative constraints for
a given event with inspiral SNR of 35 and IMR SNR of 173. The black, blue, red and dark yellow curves correspond respectively to exact likelihood, Fisher sampling (singlet-DALI), doublet-DALI and FM inversion. The FM inversion and doublet-DALI are respectively the least and the most accurate cases.
}
\label{PPN params}
\end{figure*}

\subsection{Priors and sampling}

We adopt isotropic priors for aligned spin black holes~\cite{Lange:2018pyp}, a prior uniform in comoving volume  and source frame time for the luminosity distance, and a uniform prior on component masses. The priors on extrinsic parameters are also displayed on Table~\ref{distr external params}. Since sampling is done in chirp mass and mass ratio, in Table \ref{distr external params} we display the corresponding priors on these variables, along with all other priors. In appendix \ref{app_prior} we provide further detail on the derivation of priors on chirp mass and mass ratio.

\section{Results}

\subsection{A representative example}

In Figure~\ref{PPN params} we show an example of applying \texttt{DALI} to a single event with inspiral SNR of 35 and IMR SNR of 173. Yellow denotes the posterior from FM inversion, blue the sampled Fisher, red the doublet \texttt{DALI} and black the result from sampling the exact likelihood. Despite the high total SNR we see that FM inversion still provides a poor approximation. We observe considerable improvement on the quality of the approximation by sampling the Fisher likelihood with exact priors, and further improvement by going beyond Fisher with the doublet. In order to get a statistical insight on the overall performance of the \texttt{DALI} algorithm, in the next section we display our results for a set of events and quantify the quality of the approximation with the JSD (see~\cite{deSouza:2025qok} for a discussion of the JSD in this context).
 
\subsection{Approximation performance}

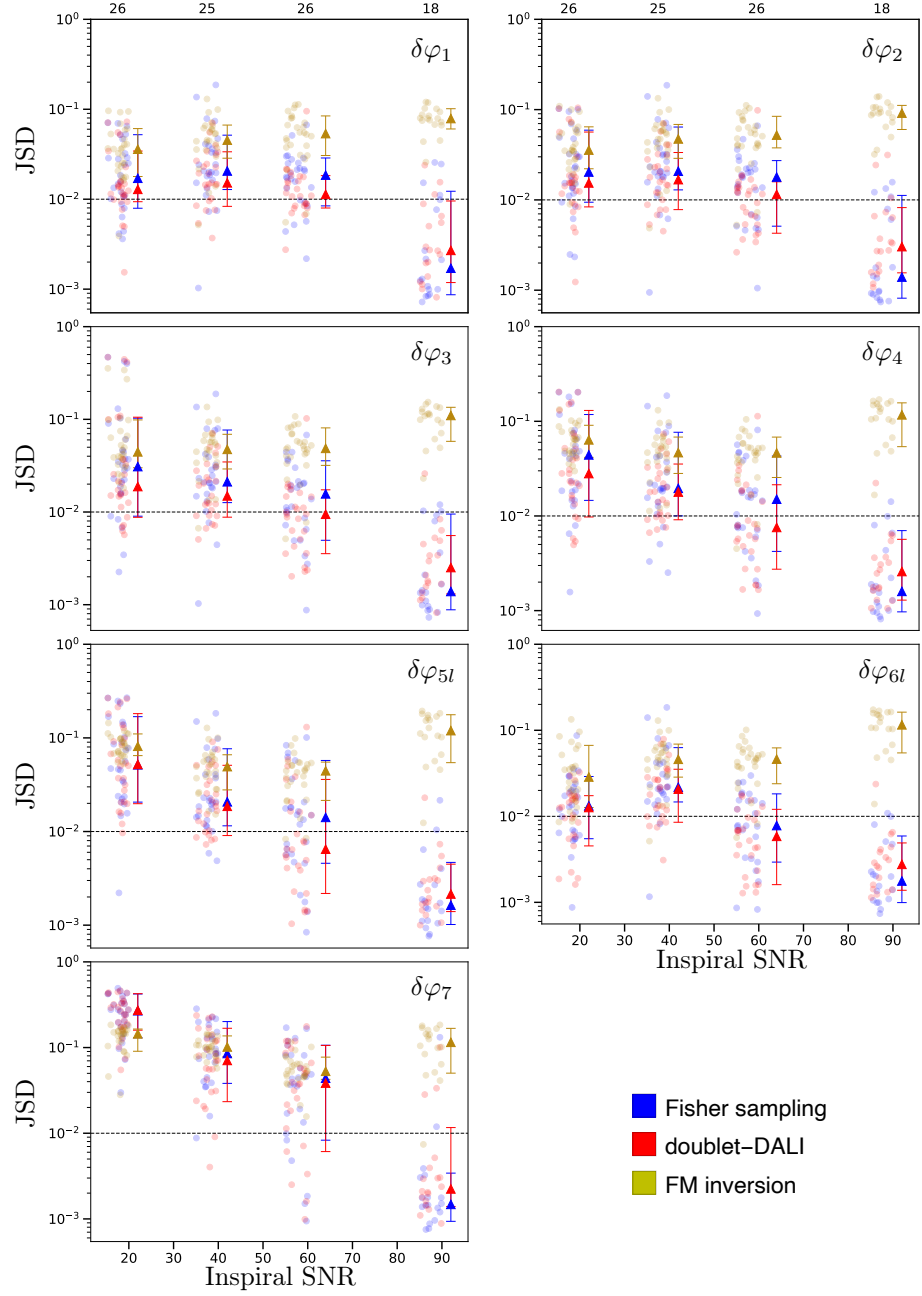
\begin{figure*}
\begin{tikzpicture}
\node[yshift=2cm, xshift=-1cm] (img0) {\includegraphics[width=0.32\textwidth]{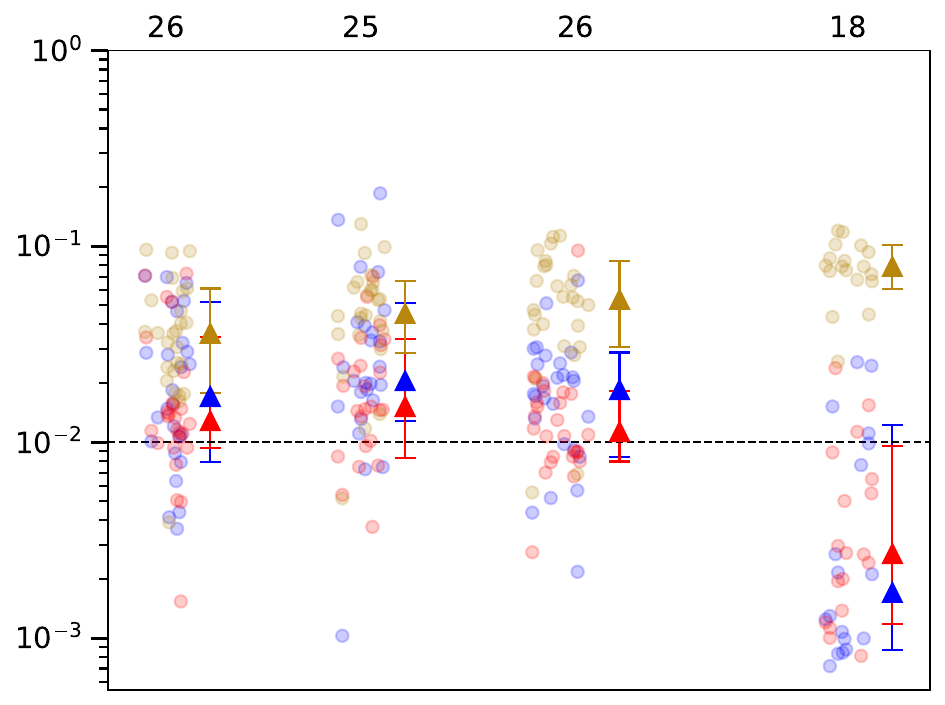}};
\node[left=of img0, node distance=0cm, rotate=90, anchor=center, yshift=-0.9cm, xshift=0.1cm, font=\fontsize{10}{10}\selectfont \color{black}] {JSD}; 
\node[above=of img0, node distance=0cm, yshift=-2.1cm, xshift=2.3cm, font=\fontsize{10}{10}\selectfont \color{black}] {$\delta \varphi_1$};

\node[right=of img0, xshift=-1cm] (img1) {\includegraphics[width=0.32\textwidth]{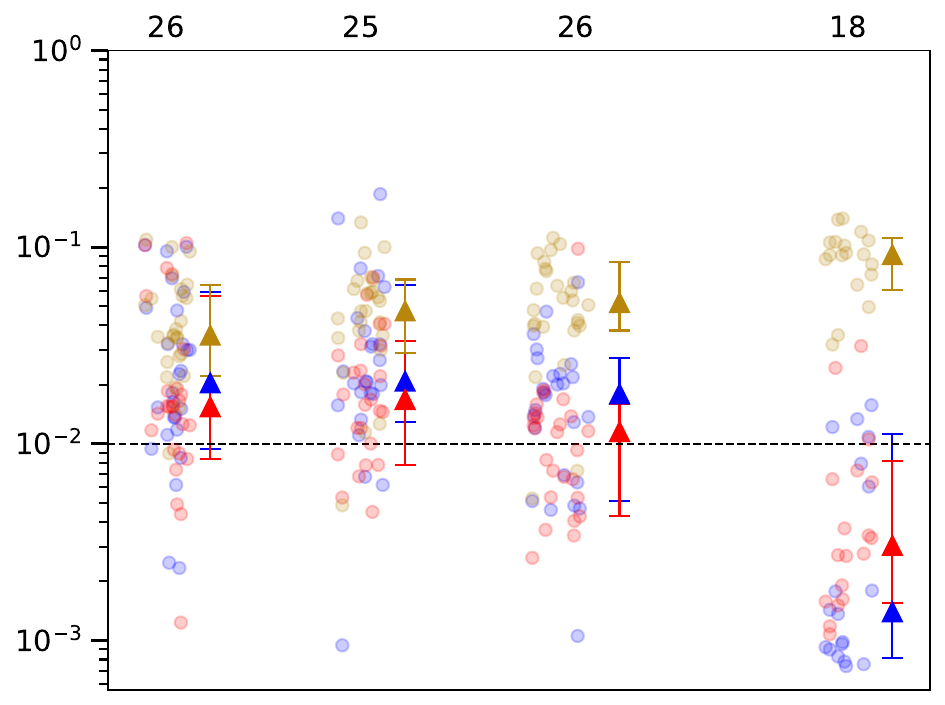}};
\node[above=of img1, node distance=0cm, yshift=-2.1cm, xshift=2.3cm, font=\fontsize{10}{10}\selectfont \color{black}] {$\delta \varphi_2$};

\node[below=of img0, yshift=1.3cm] (img2) {\includegraphics[width=0.32\textwidth]{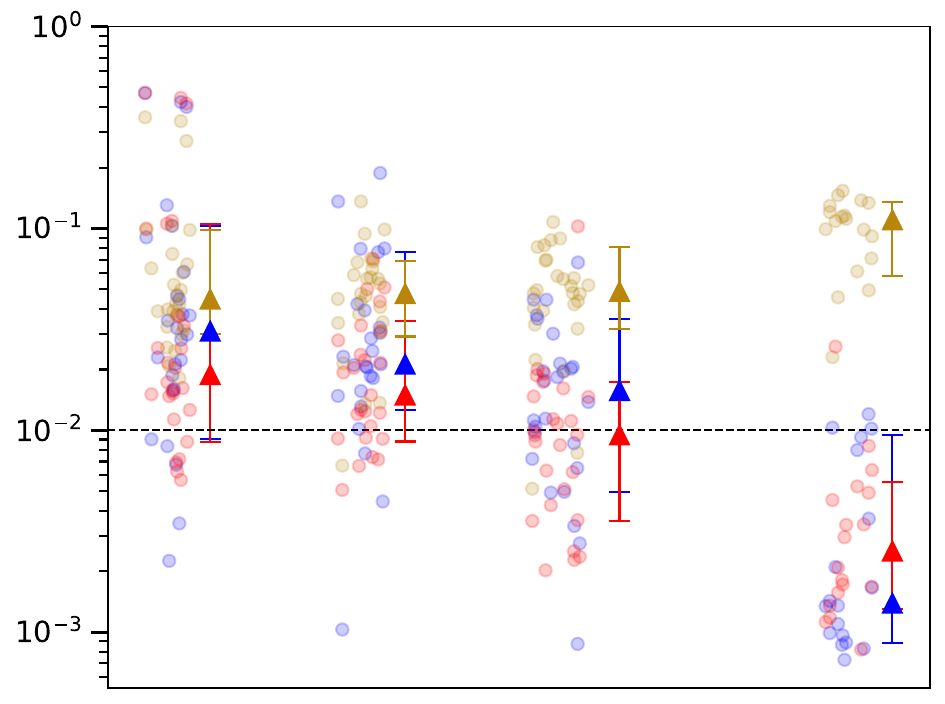}};
\node[left=of img2, node distance=0cm, rotate=90, anchor=center, yshift=-0.9cm, xshift=0.1cm, font=\fontsize{10}{10}\selectfont \color{black}] {JSD}; 
\node[above=of img2, node distance=0cm, yshift=-1.9cm, xshift=2.3cm, font=\fontsize{10}{10}\selectfont \color{black}] {$\delta \varphi_3$};

\node[below=of img1,  yshift=1.3cm] (img3) {\includegraphics[width=0.32\textwidth]{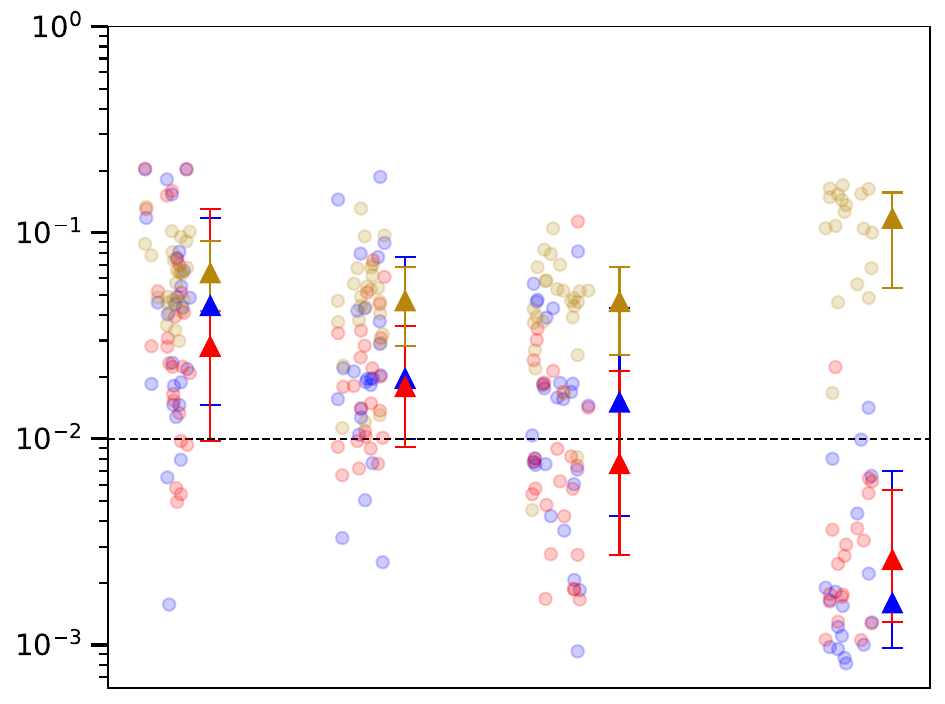}};
\node[above=of img3, node distance=0cm, yshift=-1.9cm, xshift=2.3cm, font=\fontsize{10}{10}\selectfont \color{black}] {$\delta \varphi_4$};

\node[below=of img2, yshift=1.3cm] (img4) {\includegraphics[width=0.32\textwidth]{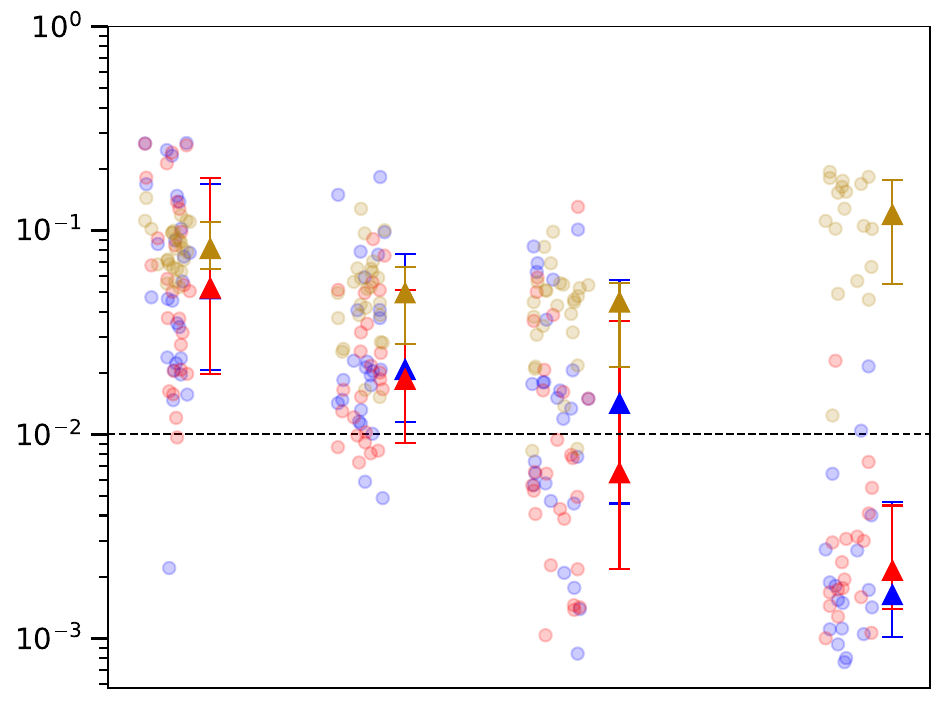}};
\node[left=of img4, node distance=0cm, rotate=90, anchor=center, yshift=-0.9cm, xshift=0.1cm, font=\fontsize{10}{10}\selectfont \color{black}] {JSD}; 
\node[above=of img4, node distance=0cm, yshift=-1.9cm, xshift=2.3cm, font=\fontsize{10}{10}\selectfont \color{black}] {$\delta \varphi_{5l}$};

\node[below=of img3, yshift=1.3cm] (img5) {\includegraphics[width=0.32\textwidth]{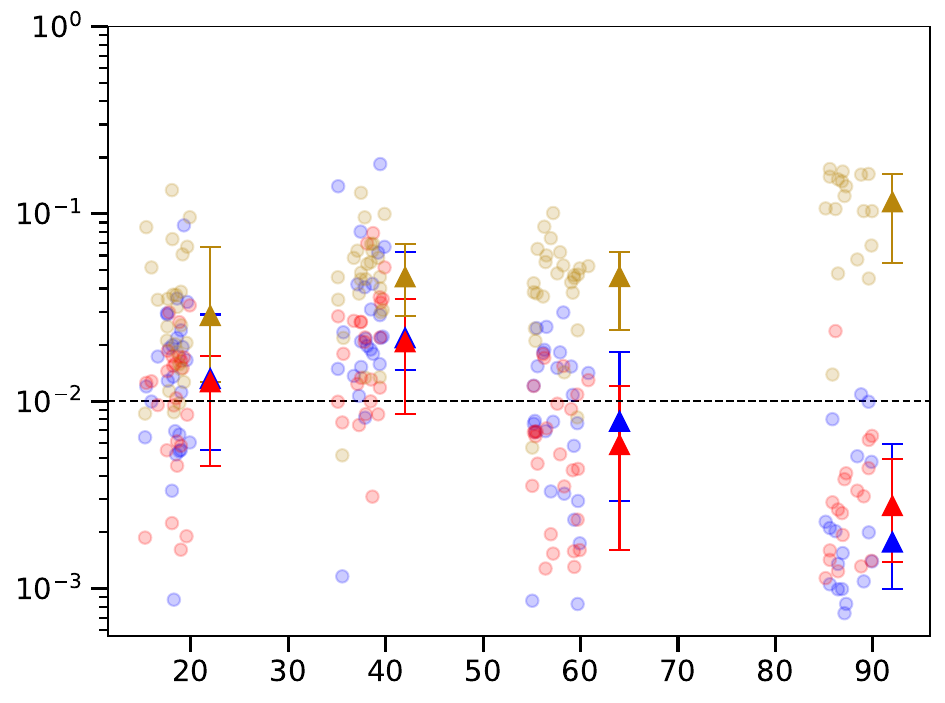}};
\node[below=of img5, node distance=0cm, yshift=1.3cm, xshift=0.3cm, font=\fontsize{10}{10}\selectfont \color{black}] {Inspiral SNR};
\node[above=of img5, node distance=0cm, yshift=-1.9cm, xshift=2.3cm, font=\fontsize{10}{10}\selectfont \color{black}] {$\delta \varphi_{6l}$};

\node[below=of img4, yshift=1.3cm] (img6) {\includegraphics[width=0.32\textwidth]{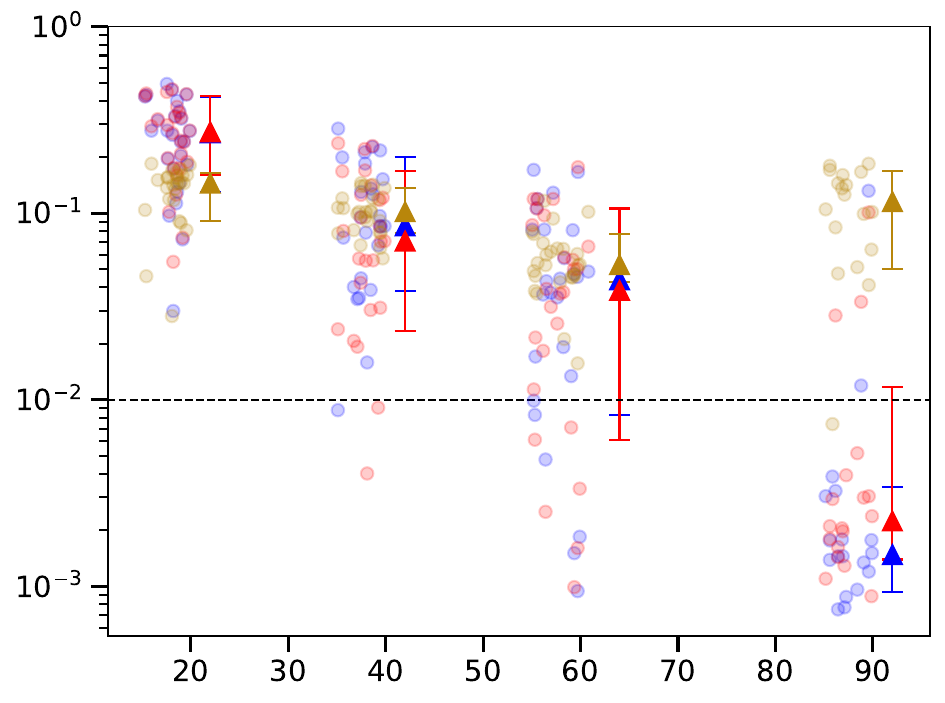}};
\node[left=of img6, node distance=0cm, rotate=90, anchor=center, yshift=-0.9cm, xshift=0.1cm, font=\fontsize{10}{10}\selectfont \color{black}] {JSD}; 
\node[below=of img6, node distance=0cm, yshift=1.3cm, xshift=0.3cm, font=\fontsize{10}{10}\selectfont \color{black}] {Inspiral SNR};
\node[above=of img6, node distance=0cm, yshift=-1.9cm, xshift=2.3cm, font=\fontsize{10}{10}\selectfont \color{black}] {$\delta \varphi_7$};
\node[below=of img5, yshift=0.8cm, xshift=0.3cm] (img6a) {\includegraphics[width=0.3\textwidth]{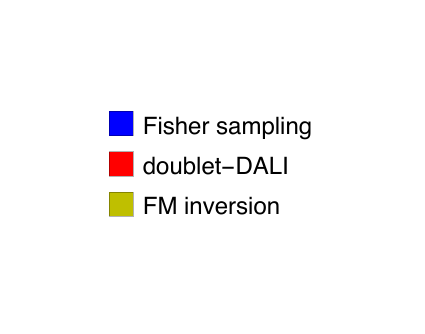}};

\end{tikzpicture}
\caption{\label{fig:JSD}  JSD for all PN parameters. Error bars denote 68\% quantiles. The dark yellow, blue and red colors denote respectively the FM, the Fisher sampling and the doublet-DALI approaches. The top horizontal axis denotes the number of events on each inspiral SNR bin (bottom horizontal axis). We see that doublet-DALI and Fisher sampling perform well, the former being better at more modest SNR, and the latter gaining an advantage at high SNR.
} 
\end{figure*}

Our main result is summarized in Figure~\ref{fig:JSD}, in which we display the JSD on the marginalized one dimensional posteriors for each of the extra GR parameters. 
The blue points correspond to the sampled Fisher (singlet), the yellow points to the FM inversion, and the red points to the doublet approximation.\footnote{For the FM inversion our approach is to extend the numerical precision of the Fisher matrix elements with \texttt{WorkingPrecision}=30 in Mathematica, and subsequently proceed with a Cholesky decomposition.} The triangles indicate the median JSD within bins of inspiral SNR centered about 20, 40, 60, and 90. The error bars represent the central 68\% interval of the JSD distribution in each bin. We also mark the line corresponding to a JSD of 0.01, below which the posteriors become qualitatively very similar. We consider that for most purposes, there is little gain in further improvements of the JSD below this threshold.

Since the FM and DALI are series expansions around the best fit for which a Gaussian is the lowest order, it is expected that these approximations will improve in accuracy as the posteriors become more Gaussian. For a fixed model, the level of Gaussianity is related to the precision  on the parameters. We note in Figure~\ref{fig:JSD} that the performance of the approximation depends weakly on the PN parameter under consideration. For example, the median JSD for $\delta\varphi_1$ using doublet is already of order $10^{-2}$ at the lowest inspiral SNR bin, whereas for $\delta\varphi_7$ it is closer to $10^{-1}$. This can be understood as an effect of the higher-order PN parameters having less impact on the GW signal, and for low SNR some degeneracies on the highest orders are still present.

A second, and more pronounced, general trend is that the JSD decreases as the inspiral SNR increases. As dicussed above, this is expected, since higher inspiral SNR signals produce posterior distributions that are more precise and thus increasingly Gaussian. The exception is the case of FM inversion, which shows a non-linear trend in JSD as a function of SNR. While further investigation is needed to pin-down the reason that the FM inversion performs relatively worse for the higher SNR bin, the most important conclusion is that in essentially all cases, the FM inversion yields the worst approximation. This further reinforces the conclusion that FM sampling with the exact priors, natural to many of the parameters involved, provides a better approximation. 

For low inspiral SNRs, there is no clear preference between sampled Fisher and the doublet approximation. Starting at an inspiral SNR of roughly 60, however, the doublet consistently yields lower JSD values, typically around $10^{-2}$, for all parameters except $\delta\varphi_7$. This marks the regime in which the inclusion of higher-order derivatives through the \texttt{DALI} expansion becomes advantageous. At even higher inspiral SNRs, around 90, the sampled Fisher and doublet approximations produces comparable results for all parameters, including $\delta\varphi_7$, which generally exhibits the least Gaussian posterior.

Considering the above, for the \texttt{IMRPhenomD} waveform model and inspiral SNR about or larger than $60$, the \texttt{DALI} algorithm provides good agreement (i.e., JSD $\lesssim 10^{-2}$) with respect to the exact likelihood (apart from $\delta\varphi_7$).
At lower inspiral SNRs, the distribution of JSD values exhibits substantial scatter, indicating that the approximation may perform well for individual events but without a high frequency of success. At very high inspiral SNRs, the posterior distributions become sufficiently Gaussian that sampled Fisher already provides an excellent approximation, reducing the practical benefit of including higher-order \texttt{DALI} corrections.

\subsection{Triplet \texttt{DALI}}

\begin{figure}[t]
\begin{tikzpicture}
\node[yshift=2cm, xshift=-1cm] (img0) {\includegraphics[width=.92\columnwidth]{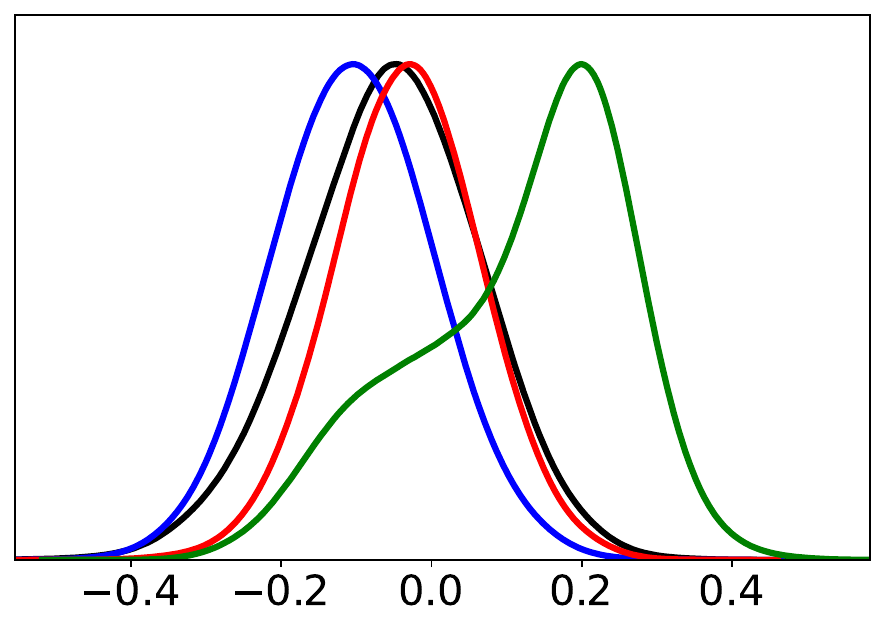}};
\node[below=of img0, node distance=0cm, yshift=1.3cm, xshift=0cm, font=\fontsize{10}{10}\selectfont \color{black}] {$\delta \varphi_1$};
\end{tikzpicture}
\caption{Black: exact likelihood. Blue: Fisher sampling. Red: doublet. Green: triplet.}
\label{triplet}
\end{figure}

Although \texttt{SymDALI} can compute the tensors of the triplet-\texttt{DALI} approximation in Eq.~\eqref{Eq:DALI}, our investigations indicate that the approximation at this order is often worse than that provided by either the Fisher sampling or the doublet. An example is shown in Fig.~\ref{triplet}. Although this behavior does not occur in every case, we found it to be frequent enough that, for \texttt{TIGER} parameters with the \texttt{IMRPhenomD} model, it is preferable to stop at the doublet approximation.


\subsection{Computational performance}

For the exact likelihood we used the nested sampling algorithm \texttt{nessai} \cite{nessai, Williams:2021qyt} through the \texttt{Bilby} plugin \cite{bilby_paper, bilby_doi} and the \texttt{Bilby} framework for signal injection. \texttt{nessai} is well known to be a good and efficient sampler in this context \cite{Santoliquido:2025aiq, Santoliquido:2025lot, Hu:2024mvn}. For the sampling of \texttt{DALI} and Fisher we used \texttt{emcee} \cite{Foreman-Mackey:2012any}, which proved to be a very efficient sampler for the \texttt{DALI} likelihood. The distinction on the chosen samplers lies on the fact that \texttt{nessai} was developed primarily for GW analysis, and it has optimizations that make the sampling of the exact likelihood more efficient, in comparison with other standard tools like \texttt{dynesty}. However, such optimized samplers are not the best choice for the \texttt{DALI} likelihood \eqref{Eq:DALI} which is computationally inexpensive, and does not benefit from complex sampling algorithms. For such likelihoods, we expect a simple MCMC algorithm to be sufficient. Regarding sampler settings, we found that for \texttt{emcee} working with 100 walkers was good enough. The amount of ensemble steps was $2.5 \times 10^5$ for the Fisher case and $5 \times 10^5$ for the doublet approximation. For \texttt{emcee}, we used the standard convergence criteria that the number of ensemble steps has to be higher than 50 autocorrelation times. The settings for \texttt{nessai} in the exact likelihood on the other hand were 5000 live points, a \texttt{volume\_fraction} of 0.99 and \texttt{reset\_flow} argument of 8. The stopping criteria for \texttt{nessai} is when the difference in the estimation of the log-evidence between sampling stages reaches 0.1. 
\begin{figure}[t]
\begin{tikzpicture}
\node[yshift=2cm, xshift=-1cm] (img0) {\includegraphics[width=.92\columnwidth]{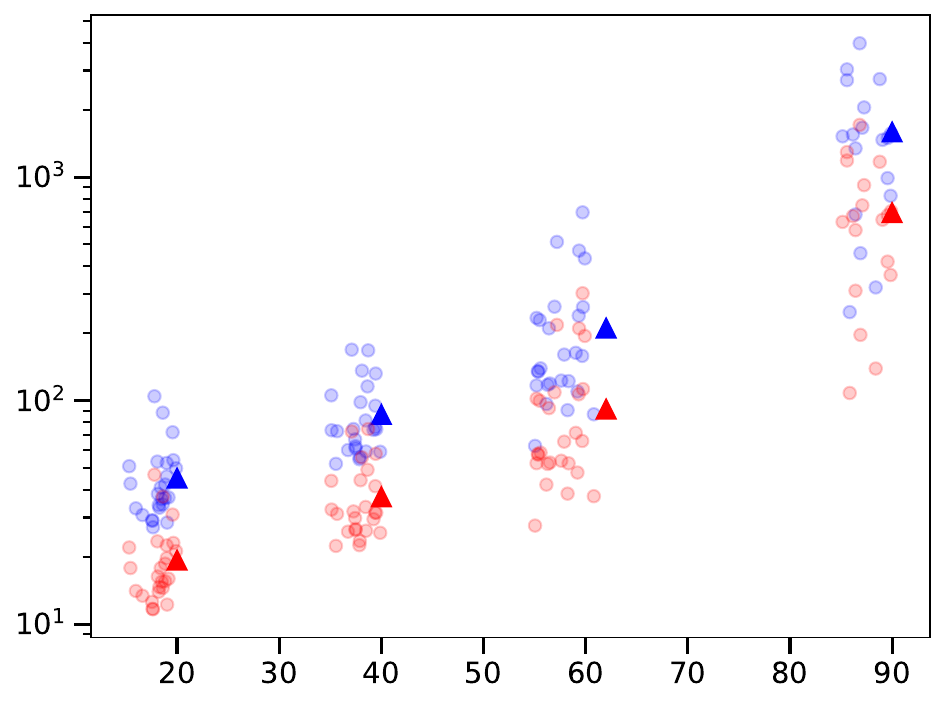}};
\node[left=of img0, node distance=0cm, rotate=90, anchor=center, yshift=-0.9cm, xshift=0.1cm, font=\fontsize{10}{10}\selectfont \color{black}] {$\Delta t_{\text{Exact}}/\Delta t_{\text{\texttt{DALI}}}$}; 
\node[below=of img0, node distance=0cm, yshift=1.3cm, xshift=0.3cm, font=\fontsize{10}{10}\selectfont \color{black}] {Inspiral SNR};
\end{tikzpicture}
\caption{Performance gain of \texttt{DALI}, considering only the time spent on the sampler, after computing all DALI tensors. Blue: sampled Fisher. Red: doublet. Circles denote the average over all $\delta \varphi_i$ per event. Triangles denote the average over events in SNR bins.}
\label{performance}
\end{figure}

A summary on the performance improvement of \texttt{DALI} with respect to the exact likelihood can be observed in Fig.~\ref{performance} which displays the ratio between the sampling time of the exact likelihood and the \texttt{DALI} algorithm as a function of inspiral SNR. We observe that the sampling time for \texttt{DALI} is constant across inspiral SNRs. The improvement in the computational benefit of \texttt{DALI} over the exact likelihood in Fig.~\ref{performance} is a consequence of the higher sampling time of the exact likelihood for high inspiral SNR. This increase derives from the fact that the time of exact likelihood calls increases with inspiral SNR,  since inspiral SNR is correlated with signal duration. Moreover, likelihoods with high inspiral SNR have posterior mass in a smaller fraction of the prior range, making them harder to sample. We do not display the time to calculate the \texttt{DALI} tensors because we found it to be inferior to 1\% of the sampling time in all cases.

\section{Other waveforms}

\begin{figure*}
\begin{tikzpicture}

\node[yshift=2cm, xshift=-1cm] (img0) {\includegraphics[width=0.49\textwidth]{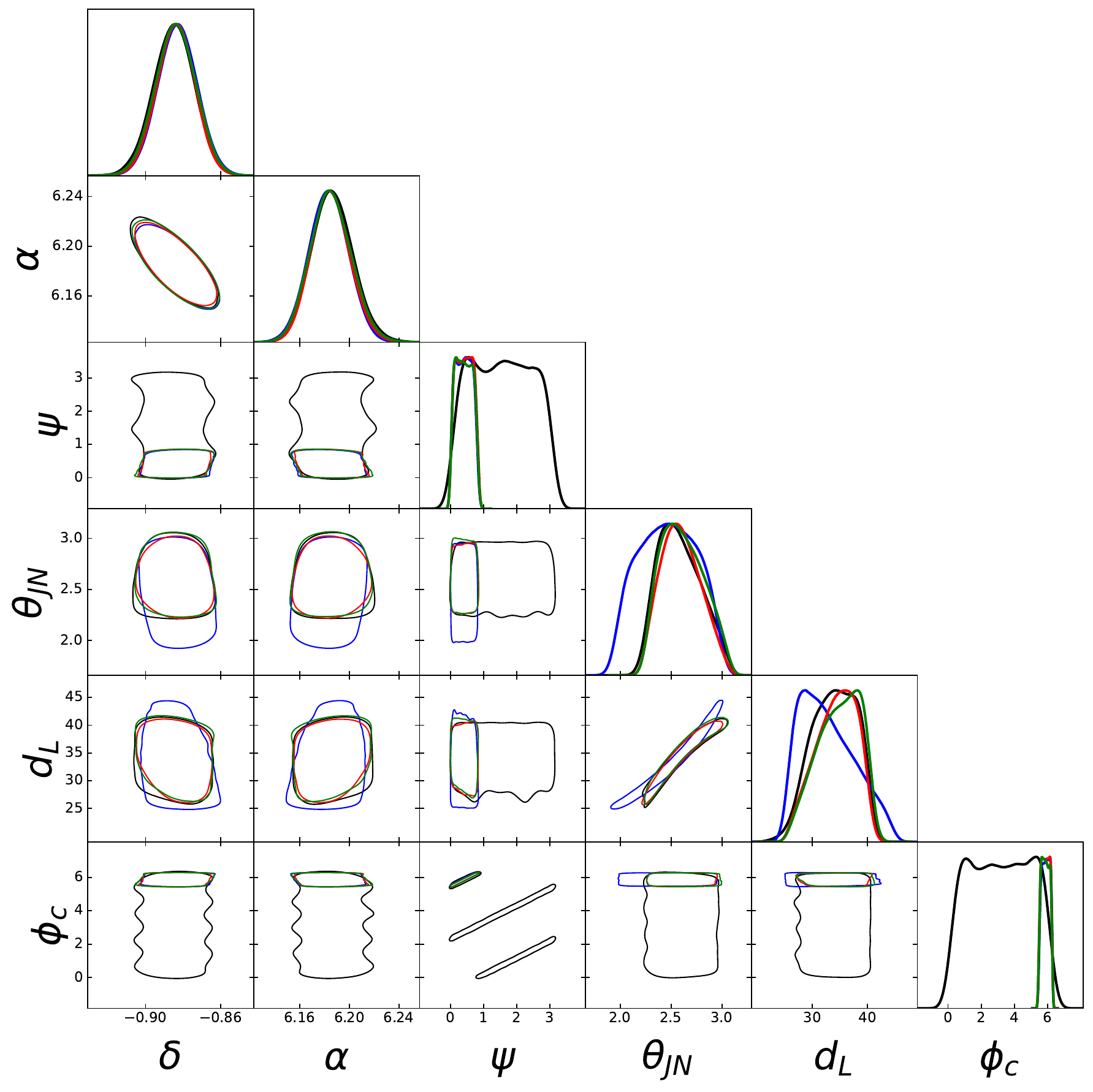}};
\node[above=of img0, node distance=0cm, yshift=-2.1cm, xshift=2.2cm, font=\fontsize{10}{10}\selectfont \color{black}] {\texttt{IMRPhenomD}};

\node[right=of img0, xshift=-1cm] (img1) {\includegraphics[width=0.49\textwidth]{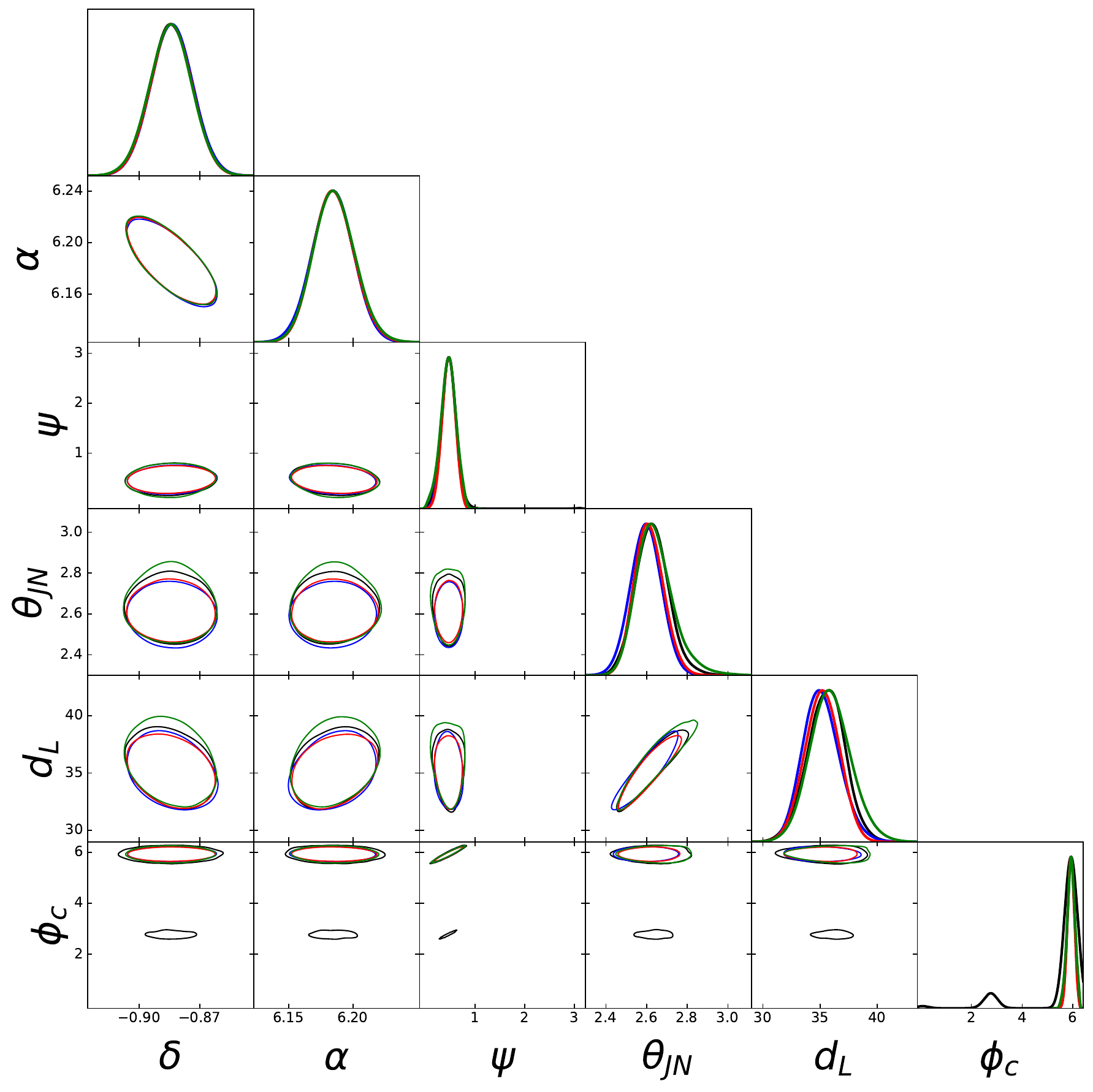}};
\node[above=of img1, node distance=0cm, yshift=-2.1cm, xshift=2.2cm, font=\fontsize{10}{10}\selectfont \color{black}] {\texttt{IMRPhenomHM}};
\end{tikzpicture}
\caption{
\label{fig:HM} Corner plot for the same event analyzed with \texttt{IMRPhenomD} (left) and \texttt{IMRPhenomHM} (right). The black, blue, red and green colors denote respectively the exact likelihood, the Fisher sampling, the doublet-DALI and triplet-DALI approaches.
} 
\end{figure*}

Although we have focused on \texttt{IMRPhenomD}, it is worth mentioning that using newer waveform models may improve the results. This applies both to the overall quality of the parameter constraints and to the performance of the approximation scheme. In particular, we found that \texttt{IMRPhenomHM} can suppress multimodalities of the posterior that are present when we use \texttt{IMRPhenomD}. In Fig.~\ref{fig:HM}, we show the exact likelihood evaluated at the same fiducial point, but analyzed using both \texttt{IMRPhenomD} and \texttt{IMRPhenomHM}. We observe that with \texttt{IMRPhenomHM}, some multimodalities of the exact likelihood disappear, leading to an overall improvement in the quality of the \texttt{DALI} approximation.

Furthermore, the computational advantage of using \texttt{DALI} with \texttt{IMRPhenomHM} is considerably greater because evaluating the exact likelihood with this waveform is more expensive than with \texttt{IMRPhenomD}. In Table \ref{tensor_calc} we display the total inference time, sampling and calculation of \texttt{DALI} tensors, for the \texttt{DALI} method in comparison with the inference time using the exact likelihood. Although \texttt{SymDALI} already supports \texttt{IMRPhenomHM} for GR parameters, it does not yet provide extended support for \texttt{TIGER} parameters. Nevertheless, we are currently investigating the viability of this scheme.

\begin{table}
\caption{Comparison on the total inference time using \texttt{DALI} and the exact likelihood with \texttt{IMRPhenomD} and \texttt{IMRPhenomHM} for the same event. Observe that Fisher denotes Fisher sampling.}
\centering
\small
\setlength{\tabcolsep}{4pt}
\label{tensor_calc}
\begin{tabular}{ccccc}
\hline
\hline
 & Fisher &  Doublet & Triplet & Exact  \\ \hline
\texttt{IMRPhenomD} &  0.11 hrs & 0.22 hrs & 0.40 hrs & 1.4 hrs \\ \hline
\texttt{IMRPhenomHM}   &  0.11 hrs & 0.23 hrs & 0.42 hrs & 69 hrs\\ 
\hline\hline
\end{tabular}
\end{table}

\section{Conclusions}

We evaluated the accuracy and computational efficiency of posterior approximations for TIGER parameters \cite{Agathos:2013upa, Roy:2025gzv} in the context of GW parameter estimation.
Our focus is on the usefulness of \texttt{DALI} to approximate GR deviations. The motivation comes from the fact that future third generation GW observatories could detect up to $10^5$ BBHs, and that full MCMC analysis of the most accurate waveforms is time consuming.

We consider posterior approximations based on Fisher matrix and the \texttt{DALI} approach \cite{Sellentin:2014zta}, whose results are compared with the exact posterior. This work continues the work \cite{deSouza:2025qok} in different ways: by developing a new code (\texttt{SymDALI}) to deal with DALI expansions symbolically, by implementing the TIGER parameters and considering \texttt{DALI} in this extended context. Similarly to \cite{deSouza:2025qok} we quantify the quality of the approximations using the Jensen–Shannon divergence (JSD).

Our main result is shown in Fig.~\ref{fig:JSD}. Using the JSD to infer the approximation performance we find that, with respect to the PN TIGER parameters: i) on average, for events with inspiral SNR about 60, the doublet DALI improves over the Fisher sampling method, which is systematically better than the standard FM inversion; ii) for lower SNR, the advantage of the Fisher sampling and doublet DALI are not evident; and iii) for higher SNR ($\sim$ 90), the Fisher sampling and doublet DALI results are essentially equivalent, while the standard FM inversion is clearly the worst approximation to the exact posterior.

About the new DALI open code, \texttt{SymDALI} \cite{SymDALI}, it includes, beyond the Fisher and DALI approximations, waveform implementation in the Wolfram Language (which can be used for other works outside the scope of this one) and beyond-GR waveform parameters (within the TIGER formalism). \texttt{SymDALI} uses the native Wolfram Language symbolic derivatives; it is also a particularly fast \texttt{DALI} implementation that includes C-compiled code for Likelihood sampling. The code is further discussed in Appendix \ref{symdali}.

Our results in this paper are restricted to \texttt{IMRPhenomD}. However, we expect the inclusion of waveform models with higher-order modes, such as \texttt{IMRPhenomHM}, to further improve the accuracy of the \texttt{DALI} approximation. More broadly, we expect \texttt{SymDALI} to be useful for forecasting GR parameters in BBH coalescences within the \texttt{DALI} framework, providing a computationally efficient tool for assessing the science case of future gravitational-wave detectors. Given the computational performance demonstrated in this work, a natural direction for future developments is to extend \texttt{SymDALI} to a broader range of waveform models.

\bigskip

\noindent
\textbf{Acknowledgments} 

FASB thanks Michele Mancarella for hospitality and useful discussions on convergence and choice of sampler for the \texttt{DALI} likelihood. FASB also thanks Alessandro Agapito for suggesting \texttt{nessai} for the exact likelihood and Andrea Begnoni for technical information on TIGER parameters.  FASB thanks Viola De Renzis for suggestions on prior choice and likelihood marginalization; Lijing Shao and Ziming Wang for technical details on the waveform implementation and Michael Williams for suggestions on the settings of  \texttt{nessai}. FASB also thanks the active \texttt{ripple} community \cite{Edwards:2023sak}, whose open source code was referenced to implement waveforms. FASB acknowledges support from \textit{Fundação de Amparo à Pesquisa e Inovação do Espírito Santo} (FAPES-Brazil). DCR acknowledges \textit{Conselho Nacional de Desenvolvimento Científico e Tecnológico} (CNPq-Brazil) and FAPES for partial support. JMSdS acknowledges support by FAPES and \textit{Fundação Carlos Chagas Filho de Amparo à Pesquisa do Estado do Rio de Janeiro} (FAPERJ), project E-26/200.236/2024 and E-26/200.237/2024. MQ is supported by the Brazilian research agencies FAPERJ project E-26/201.237/2022, CNPq (Conselho Nacional de Desenvolvimento Científico e Tecnológico) and CAPES. 
The authors acknowledge the use of computational resources from the \texttt{Sci-Com} Lab of the Physics Department at UFES, supported by FAPES, CAPES, and CNPq.


\appendix

\bigskip


\setlength{\itemsep}{-5pt}
\setlength{\parsep}{0pt}
\setlength{\topsep}{2pt}
\renewcommand{\labelitemi}{--}

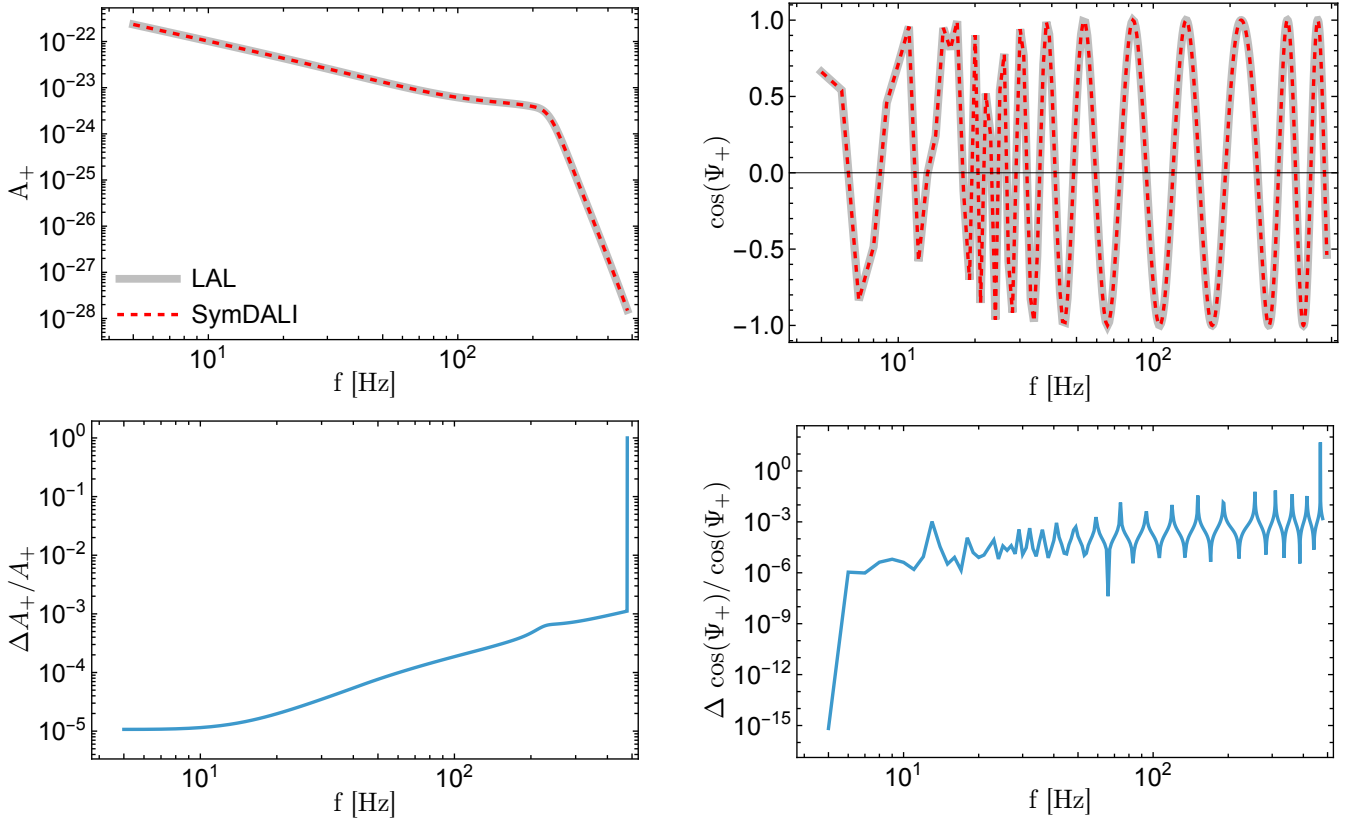
\begin{figure*}
\begin{tikzpicture}
\node[yshift=2cm, xshift=-1cm] (img0) {\includegraphics[width=0.45\textwidth]{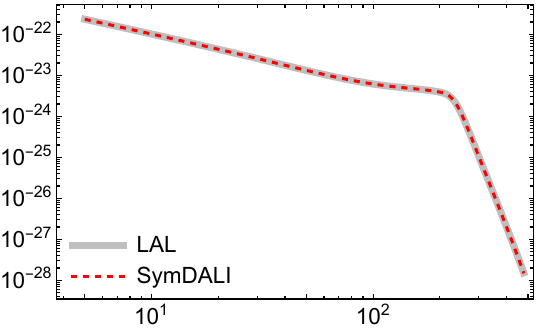}};
\node[left=of img0, node distance=0cm, rotate=90, anchor=center, yshift=-0.9cm, xshift=0.1cm, font=\fontsize{10}{10}\selectfont \color{black}] {A$_+$}; 
\node[below=of img0, node distance=0cm, yshift=1.3cm, xshift=0.3cm, font=\fontsize{10}{10}\selectfont \color{black}] {f [Hz]};

\node[right=of img0, xshift=-0.1cm] (img1) {\includegraphics[width=0.45\textwidth]{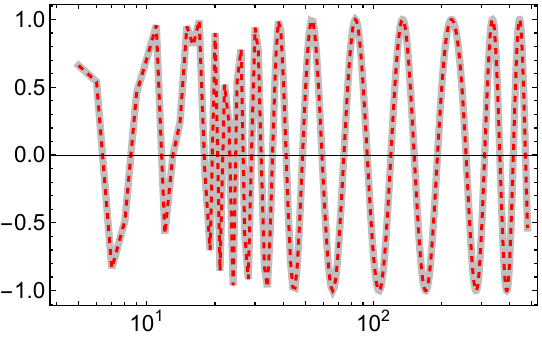}};
\node[left=of img1, node distance=0cm, rotate=90, anchor=center, yshift=-0.9cm, xshift=0.1cm, font=\fontsize{10}{10}\selectfont \color{black}] {$\cos(\Psi_+)$}; 
\node[below=of img1, node distance=0cm, yshift=1.3cm, xshift=0.3cm, font=\fontsize{10}{10}\selectfont \color{black}] {f [Hz]};

\node[below=of img0, yshift=0.7cm] (img2) {\includegraphics[width=0.45\textwidth]{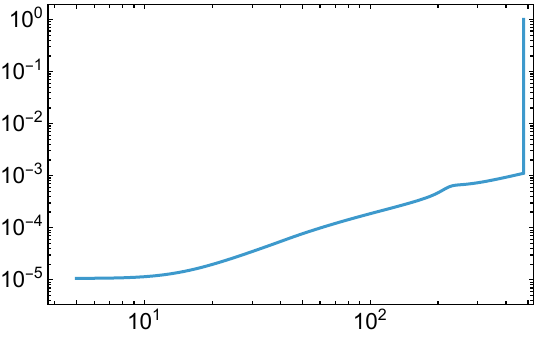}};
\node[left=of img2, node distance=0cm, rotate=90, anchor=center, yshift=-0.9cm, xshift=0.1cm, font=\fontsize{10}{10}\selectfont \color{black}] {$\Delta A_+/A_+$}; 
\node[below=of img2, node distance=0cm, yshift=1.3cm, xshift=0.3cm, font=\fontsize{10}{10}\selectfont \color{black}] {f [Hz]};

\node[below=of img1,  yshift=0.7cm] (img3) {\includegraphics[width=0.45\textwidth]{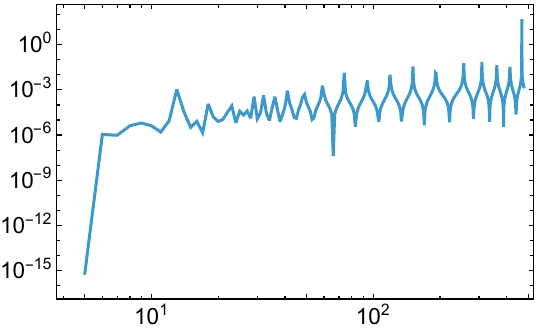}};
\node[left=of img3, node distance=0cm, rotate=90, anchor=center, yshift=-0.9cm, xshift=0.1cm, font=\fontsize{10}{10}\selectfont \color{black}] {$\Delta \, \cos(\Psi_+)/\cos(\Psi_+)$}; 

\node[below=of img3, node distance=0cm, yshift=1.3cm, xshift=0.3cm, font=\fontsize{10}{10}\selectfont \color{black}] {f [Hz]};

\end{tikzpicture}
\caption{\label{fig:WF_comp} Comparison between \texttt{LAL} (gray) and \texttt{SymDALI} (red) implementations of \texttt{IMRPhenomD}. We observe that the peak at the end of the amplitude residual is due to the cuttof frequency of the waveform, where the \texttt{LAL} results is zero for the last frequency and \texttt{SymDALI} is not. The chosen parameters are $\mathcal{M}_c=36 \, M_\odot$, $\eta=0.24$, $\chi_1  = 0.8$, $\chi_2=-0.8$, $d_L = 0.98$ Gpc and $\theta_{JN} = \phi_c = 0$.
} 
\end{figure*}

\section{Waveform implementation}\label{app_comp}

In Fig.~\ref{fig:WF_comp} we provide a comparison between the implementations of \texttt{IMRPhenomD} in \texttt{LAL} and in \texttt{SymDALI}. The residuals between implementations are calculated as the relative difference between both implementations. We choose the amplitude and cosine of the plus polarization.

\section{Chirp mass and mass ratio prior}\label{app_prior}


Here we derive the priors on the chirp mass and mass ratio assuming uniform priors on the component masses. Given the probability distribution transformation law
\begin{align}
    \tilde{\pi}(\boldsymbol{y}) = \pi(\boldsymbol{x}(\boldsymbol{y})) \; J,
\end{align}
where $J$ is the Jacobian $J \equiv | \partial \, \boldsymbol{x}(\boldsymbol{y})/  \partial \boldsymbol{y}|$ and assuming that the prior on $(m_1, m_2)$ is uniform we have
that the prior on $(\mathcal{M}_c, \, q)$ is proportional to the Jacobian of the transformation. Making use of the definitions
\begin{align*}
    &M = m_1+m_2,\\
    &\eta = \frac{m_1 m_2}{M^2},\\
    &\mathcal{M}_c = M \, \eta^{3/5},\\
    & q = \frac{m_2}{m_1}.
\end{align*}
we have 
\begin{align}
    \pi(\mathcal{M}_c, \, q) \propto J = \mathcal{M}_c  \, \frac{(1+q)^{2/5}}{q^{6/5}}.
\end{align}
The prior on $(m_1, m_2)$ is defined for $m_1 \geq m_2$ and $m_1 \in (m_1^\text{min}, \; m_1^\text{max}), \, m_2 \in (m_1^\text{min}, \, m_1^\text{max})$. 

As a consequence in the $(\mathcal{M}_c, \, q)$ plane the prior is defined for $q \in (m_1^\text{min}/ m_1^\text{max}, \, 1)$ and 
\begin{align}
    \mathcal{M}_c \in \left[ 
    \frac{m_1^\text{min}}{(1+q)^{1/5} \, q^{2/5}}, \; \frac{m_1^\text{max} \, q^{3/5}}{(1+q)^{1/5}}
    \right].
\end{align}
In practice, since we are sampling directly in $(\mathcal{M}_c, \, q)$ it is more practical to define independent intervals for $(\mathcal{M}_c, \, q)$. 
As long as the posterior mass at the boundaries is negligible, there will be a corresponding interval for $m_1$ such that in the $(\mathcal{M}_c, \, q)$ plane we have enough coverage for the posterior.  Here we take $q \in [0.125, \, 1]$ 
and $\mathcal{M}_c \in \left[ \mathcal{M}_c^{\text{fid}} - 50 \, M_\odot, \, \mathcal{M}_c^{\text{fid}} +50 \, M_\odot\right].$\footnote{There are six events in our set for which such bound allows for negative chirp masses. However, we checked by inspection that none of these events produced posterior samples close to $\mathcal{M}_c = 0$, meaning that running the parameter estimation again with appropriate prior bounds is not going to affect the results.} We found these intervals to be sufficiently broad that the exact posterior remains well within the prior bounds.

\section{\texttt{SymDALI}}\label{symdali}

It is not possible to implement straightforward derivatives of waveforms in Wolfram Language, since these are complex functions. For this work we developed a structure similar to autodiff, which is limited, problem oriented and does not compare to tools like \texttt{jax}. The framework is based on two functions we created: \texttt{\$Block} and \texttt{DerivativeRules}. \texttt{\$Block} works similarly to the built in \texttt{Block} symbol, however when one declares:

\begin{mmaCell}[moredefined=DerivativeRules,moredefined=$Block]{Code}
DerivativeRules[
    {Test, {x}, 2},
    $Block[
        {{\mmaFnc{h1}, \mmaFnc{h2}}},
        \mmaFnc{h1}[\mmaPat{y_}] := Sin[\mmaPat{y}];
        \mmaFnc{h2}[\mmaPat{y_}] := \mmaPat{y}^2;
        \mmaFnc{h2}[\mmaFnc{h1}[\mmaPat{x}]]
    ]
]
\end{mmaCell}
we get
\begin{mmaCell}{Output}
\{Test\textquotesingle[x] -> Block[
    \{\mmaFnc{x1}, \mmaFnc{x2}, \mmaFnc{x3}\},
    \mmaFnc{x1} = Sin[x];
    \mmaFnc{x2} = Cos[x];
    \mmaFnc{x3} = 2 \mmaFnc{x1};
    \mmaFnc{x2} \mmaFnc{x3} 
],
Test\textquotesingle\textquotesingle[x] -> Block[
    \{\mmaFnc{x1}, \mmaFnc{x2}, \mmaFnc{x3}, \mmaFnc{x4}\},
    \mmaFnc{x1} = Sin[x];
    \mmaFnc{x2} = Cos[x];
    \mmaFnc{x3} = 2 \mmaFnc{x1};
    \mmaFnc{x4} = -Sin[x];
    2 \mmaFnc{x2}^2  + \mmaFnc{x3} \mmaFnc{x4}
]\}
\end{mmaCell}
The output is the first two derivatives of the original function. Higher-order derivatives are obtained by replacing \texttt{2} in \texttt{DerivativeRules} with the desired order n. The list \texttt{{x}} specifies the variables with respect to which derivatives are calculated. The intermediate variables \texttt{{h1, h2}} inside \texttt{\$Block} are tracked by \texttt{DerivativeRules}, allowing the derivative of the complete function to be constructed from these intermediate functions and their derivatives. Since this is a framework developed for \texttt{SymDALI} we perform tests to make sure that the correct derivatives of waveforms have been calculated.

The symbolic expressions produced by \texttt{DerivativeRules} are common Wolfram Language expressions that are evaluated by the kernel. We can get extra performance by using \texttt{Compile} which generates expressions that are evaluated by the Wolfram virtual machine (WVM). These compiled functions, representing the waveform and its derivatives, can be saved to disk and called at run time.  

When the user starts \texttt{SymDALI} all necessary derivatives of the waveform are saved in the disk in the form of compiled WVM functions. At run time \texttt{SymDALI} will basically call the correct functions and evaluate them at the desired input values.

Once the tensors are calculated, Eq.~\eqref{Eq:DALI} still needs to be sampled so we can estimate the posterior. In this stage \texttt{SymDALI} makes use of the symmetries in the problem to speed up likelihood calls. We observe that when contracting an arbitrary tensor $T_{i_1 \dots i_N}$ with a tensor defined by $\Delta \theta^{i_1} \dots \Delta \theta^{i_N}$, only the totally symmetric part of $T_{i_1 \dots i_N}$ contributes to the result. Moreover, the contraction of totally symmetric tensors carries a lot of redundancies. For instance, given the symmetric matrix $M_{ij}$ we have
\begin{align*}
    M_{ij} \, x_i x_j  &= M_{11} \, (x_1)^2  +  M_{12} \, x_1 x_2 + M_{12} \, x_1 x_2 + M_{22} (x_2)^2\\
    &=  M_{11} \, (x_1)^2  +  2 M_{12} \, x_1 x_2 +  M_{22} (x_2)^2,\nonumber
\end{align*}
where we used $M_{12}=M_{21}$. This redundancy can be eliminated by defining $N_{ij}$, with $j \ge i$, such that $N_{ii}=M_{ii}$ and $N_{12}=2M_{12}$, such that 
\begin{align*}
     M_{ij} \, x_i x_j = \sum_{j \geq i} N_{ij} \, x_i x_j.
\end{align*}
This can be extended to rank-N symmetric tensors, such that we can perform the contractions in Eq.~\eqref{Eq:DALI} constraining the sum to $i_1\le\cdots\le i_N$ with the appropriate $N_{i_1 \dots i_N}$ objects. In addition to that, the calculation of the powers $\Delta \theta^{i_1 \dots i_N}$ themselves is done in compiled C code, accessed through numpy, which gives an extra performance boost. This strategy results in fast likelihood calls in \texttt{SymDALI}.

\bibliographystyle{apsrev4-1}
\bibliography{sample} 

@article{Tagliazucchi:2026dpr,
    author = "Tagliazucchi, Matteo and Moresco, Michele and Agapito, Alessandro and Mancarella, Michele and Ferraiuolo, Sarah and Mastrogiovanni, Simone and Borghi, Nicola and Pannarale, Francesco and Bonacorsi, Daniele",
    title = "{Pushing spectral siren cosmology into the third-generation era: a blinded mock data challenge}",
    eprint = "2602.17756",
    archivePrefix = "arXiv",
    primaryClass = "astro-ph.CO",
    month = "2",
    year = "2026"
}

@article{Cutler:1994ys,
	archiveprefix = {arXiv},
	author = {Cutler, Curt and Flanagan, Eanna E.},
	bdsk-color = {4},
	doi = {10.1103/PhysRevD.49.2658},
	eprint = {gr-qc/9402014},
	journal = {Phys. Rev. D},
	pages = {2658--2697},
	reportnumber = {GRP-369},
	title = {{Gravitational waves from merging compact binaries: How accurately can one extract the binary's parameters from the inspiral wave form?}},
	volume = {49},
	year = {1994}
}

@article{Finn:1992wt,
	archiveprefix = {arXiv},
	author = {Finn, Lee S.},
	doi = {10.1103/PhysRevD.46.5236},
	eprint = {gr-qc/9209010},
	journal = {Phys. Rev. D},
	pages = {5236--5249},
	reportnumber = {PRINT-93-0128 (NORTHWESTERN)},
	title = {{Detection, measurement and gravitational radiation}},
	volume = {46},
	year = {1992}
}

@article{EfronHinkley1978,
  author  = {Efron, Bradley and Hinkley, David V.},
  title   = {Assessing the Accuracy of the Maximum Likelihood Estimator:
             Observed Versus Expected Fisher Information},
  journal = {Biometrika},
  year    = {1978},
  volume  = {65},
  number  = {3},
  pages   = {457--483},
  doi     = {10.1093/biomet/65.3.457}
}

@article{autodiff1964,
author = {Wengert, R. E.},
title = {A simple automatic derivative evaluation program},
year = {1964},
issue_date = {Aug. 1964},
publisher = {Association for Computing Machinery},
address = {New York, NY, USA},
volume = {7},
number = {8},
issn = {0001-0782},
url = {https://doi.org/10.1145/355586.364791},
doi = {10.1145/355586.364791},
journal = {Commun. ACM},
month = aug,
pages = {463–464},
numpages = {2}
}

@article{wang:2022kia,
  author	= "Wang, Ziming and Liu, Chang and Zhao, Junjie and Shao,
		  Lijing",
  title		= "{Extending the Fisher Information Matrix in
		  Gravitational-wave Data Analysis}",
  eprint	= "2203.02670",
  archiveprefix	= "arXiv",
  primaryclass	= "gr-qc",
  doi		= "10.3847/1538-4357/ac6b99",
  journal	= "Astrophys. J.",
  volume	= "932",
  number	= "2",
  pages		= "102",
  year		= "2022"
}

@article{Lange:2018pyp,
    author = "Lange, Jacob and O'Shaughnessy, Richard and Rizzo, Monica",
    title = "{Rapid and accurate parameter inference for coalescing, precessing compact binaries}",
    eprint = "1805.10457",
    archivePrefix = "arXiv",
    primaryClass = "gr-qc",
    reportNumber = "LIGO DCC P1800084, LIGO-DCC-P1800084",
    month = "5",
    year = "2018"
}

@article{desouza:2023ozp,
  author	= "de Souza, Josiel Mendon\c{c}a Soares and Sturani,
		  Riccardo",
  title		= "{GWDALI: A Fisher-matrix based software for gravitational
		  wave parameter-estimation beyond Gaussian approximation}",
  eprint	= "2307.10154",
  archiveprefix	= "arXiv",
  primaryclass	= "gr-qc",
  doi		= "10.1016/j.ascom.2023.100759",
  journal	= "Astron. Comput.",
  volume	= "45",
  pages		= "100759",
  year		= "2023"
}

@article{deSouza:2025qok,
    author = "de Souza, Josiel Mendon{\c{c}}a Soares and Quartin, Miguel",
    title = "{On the use of the Derivative Approximation for Likelihoods for gravitational wave inference}",
    eprint = "2510.16955",
    archivePrefix = "arXiv",
    primaryClass = "astro-ph.IM",
    doi = "10.1088/1475-7516/2026/05/101",
    journal = "JCAP",
    volume = "05",
    pages = "101",
    year = "2026"
}

@article{Dupletsa:2022scg,
    author = "Dupletsa, Ulyana and Harms, Jan and Banerjee, Biswajit and Branchesi, Marica and Goncharov, Boris and Maselli, Andrea and Oliveira, Ana Carolina Silva and Ronchini, Samuele and Tissino, Jacopo",
    title = "{gwfish: A simulation software to evaluate parameter-estimation capabilities of gravitational-wave detector networks}",
    eprint = "2205.02499",
    archivePrefix = "arXiv",
    primaryClass = "gr-qc",
    doi = "10.1016/j.ascom.2022.100671",
    journal = "Astron. Comput.",
    volume = "42",
    pages = "100671",
    year = "2023"
}

@article{Borhanian:2020ypi,
    author = "Borhanian, Ssohrab",
    title = "{GWBENCH: a novel Fisher information package for gravitational-wave benchmarking}",
    eprint = "2010.15202",
    archivePrefix = "arXiv",
    primaryClass = "gr-qc",
    doi = "10.1088/1361-6382/ac1618",
    journal = "Class. Quant. Grav.",
    volume = "38",
    number = "17",
    pages = "175014",
    year = "2021"
}

@article{Iacovelli:2022mbg,
    author = "Iacovelli, Francesco and Mancarella, Michele and Foffa, Stefano and Maggiore, Michele",
    title = "{GWFAST: A Fisher Information Matrix Python Code for Third-generation Gravitational-wave Detectors}",
    eprint = "2207.06910",
    archivePrefix = "arXiv",
    primaryClass = "astro-ph.IM",
    doi = "10.3847/1538-4365/ac9129",
    journal = "Astrophys. J. Supp.",
    volume = "263",
    number = "1",
    pages = "2",
    year = "2022"
}

@article{Tegmark:1997rp,
    author = "Tegmark, Max",
    title = "{Measuring cosmological parameters with galaxy surveys}",
    eprint = "astro-ph/9706198",
    archivePrefix = "arXiv",
    reportNumber = "IASSNS-AST-97-44",
    doi = "10.1103/PhysRevLett.79.3806",
    journal = "Phys. Rev. Lett.",
    volume = "79",
    pages = "3806--3809",
    year = "1997"
}

@article{Roy:2025gzv,
    author = "Roy, Soumen and Haney, Maria and Pratten, Geraint and T. H. Pang, Peter and Van Den Broeck, Chris",
    title = "{Improved parametrized test of general relativity using the IMRPhenomX waveform family: Including higher harmonics and precession}",
    eprint = "2504.21147",
    archivePrefix = "arXiv",
    primaryClass = "gr-qc",
    reportNumber = "LIGO DCC P2500034",
    doi = "10.1103/855k-sys5",
    journal = "Phys. Rev. D",
    volume = "113",
    number = "2",
    pages = "024016",
    year = "2026"
}

@article{Punturo:2010zz,
    author = "Punturo, M. and others",
    editor = "Ricci, Fulvio",
    title = "{The Einstein Telescope: A third-generation gravitational wave observatory}",
    doi = "10.1088/0264-9381/27/19/194002",
    journal = "Class. Quant. Grav.",
    volume = "27",
    pages = "194002",
    year = "2010"
}

@article{Williams:2021qyt,
    author = "Williams, Michael J. and Veitch, John and Messenger, Chris",
    title = "{Nested sampling with normalizing flows for gravitational-wave inference}",
    eprint = "2102.11056",
    archivePrefix = "arXiv",
    primaryClass = "gr-qc",
    doi = "10.1103/PhysRevD.103.103006",
    journal = "Phys. Rev. D",
    volume = "103",
    number = "10",
    pages = "103006",
    year = "2021"
}

@article{Branchesi:2023mws,
    author = "Branchesi, Marica and others",
    title = "{Science with the Einstein Telescope: a comparison of different designs}",
    eprint = "2303.15923",
    archivePrefix = "arXiv",
    primaryClass = "gr-qc",
    reportNumber = "ET-0084A-23",
    doi = "10.1088/1475-7516/2023/07/068",
    journal = "JCAP",
    volume = "07",
    pages = "068",
    year = "2023"
}

@article{Planck:2018vyg,
    author = "Aghanim, N. and others",
    collaboration = "Planck",
    title = "{Planck 2018 results. VI. Cosmological parameters}",
    eprint = "1807.06209",
    archivePrefix = "arXiv",
    primaryClass = "astro-ph.CO",
    doi = "10.1051/0004-6361/201833910",
    journal = "Astron. Astrophys.",
    volume = "641",
    pages = "A6",
    year = "2020",
    note = "[Erratum: Astron.Astrophys. 652, C4 (2021)]"
}

@article{Khan:2015jqa,
    author = "Khan, Sebastian and Husa, Sascha and Hannam, Mark and Ohme, Frank and P{\"u}rrer, Michael and Jim{\'e}nez Forteza, Xisco and Boh{\'e}, Alejandro",
    title = "{Frequency-domain gravitational waves from nonprecessing black-hole binaries. II. A phenomenological model for the advanced detector era}",
    eprint = "1508.07253",
    archivePrefix = "arXiv",
    primaryClass = "gr-qc",
    doi = "10.1103/PhysRevD.93.044007",
    journal = "Phys. Rev. D",
    volume = "93",
    number = "4",
    pages = "044007",
    year = "2016"
}

@article{Husa:2015iqa,
    author = "Husa, Sascha and Khan, Sebastian and Hannam, Mark and P{\"u}rrer, Michael and Ohme, Frank and Jim{\'e}nez Forteza, Xisco and Boh{\'e}, Alejandro",
    title = "{Frequency-domain gravitational waves from nonprecessing black-hole binaries. I. New numerical waveforms and anatomy of the signal}",
    eprint = "1508.07250",
    archivePrefix = "arXiv",
    primaryClass = "gr-qc",
    doi = "10.1103/PhysRevD.93.044006",
    journal = "Phys. Rev. D",
    volume = "93",
    number = "4",
    pages = "044006",
    year = "2016"
}

@article{LIGOScientific:2019fpa,
    author = "Abbott, B. P. and others",
    collaboration = "LIGO Scientific, Virgo",
    title = "{Tests of General Relativity with the Binary Black Hole Signals from the LIGO-Virgo Catalog GWTC-1}",
    eprint = "1903.04467",
    archivePrefix = "arXiv",
    primaryClass = "gr-qc",
    reportNumber = "LIGO-P1800316",
    doi = "10.1103/PhysRevD.100.104036",
    journal = "Phys. Rev. D",
    volume = "100",
    number = "10",
    pages = "104036",
    year = "2019"
}

@article{LIGOScientific:2020tif,
    author = "Abbott, R. and others",
    collaboration = "LIGO Scientific, Virgo",
    title = "{Tests of general relativity with binary black holes from the second LIGO-Virgo gravitational-wave transient catalog}",
    eprint = "2010.14529",
    archivePrefix = "arXiv",
    primaryClass = "gr-qc",
    reportNumber = "LIGO-P2000091",
    doi = "10.1103/PhysRevD.103.122002",
    journal = "Phys. Rev. D",
    volume = "103",
    number = "12",
    pages = "122002",
    year = "2021"
}

@article{Agathos:2013upa,
    author = "Agathos, Michalis and Del Pozzo, Walter and Li, Tjonnie G. F. and Van Den Broeck, Chris and Veitch, John and Vitale, Salvatore",
    title = "{TIGER: A data analysis pipeline for testing the strong-field dynamics of general relativity with gravitational wave signals from coalescing compact binaries}",
    eprint = "1311.0420",
    archivePrefix = "arXiv",
    primaryClass = "gr-qc",
    doi = "10.1103/PhysRevD.89.082001",
    journal = "Phys. Rev. D",
    volume = "89",
    number = "8",
    pages = "082001",
    year = "2014"
}

@article{Dupletsa:2024gfl,
    author = "Dupletsa, Ulyana and Harms, Jan and Ng, Ken K. Y. and Tissino, Jacopo and Santoliquido, Filippo and Cozzumbo, Andrea",
    title = "{Validating prior-informed Fisher-matrix analyses against GWTC data}",
    eprint = "2404.16103",
    archivePrefix = "arXiv",
    primaryClass = "gr-qc",
    doi = "10.1103/PhysRevD.111.024036",
    journal = "Phys. Rev. D",
    volume = "111",
    number = "2",
    pages = "024036",
    year = "2025"
}

@article{Edwards:2023sak,
    author = "Edwards, Thomas D. P. and Wong, Kaze W. K. and Lam, Kelvin K. H. and Coogan, Adam and Foreman-Mackey, Daniel and Isi, Maximiliano and Zimmerman, Aaron",
    title = "{Differentiable and hardware-accelerated waveforms for gravitational wave data analysis}",
    eprint = "2302.05329",
    archivePrefix = "arXiv",
    primaryClass = "astro-ph.IM",
    doi = "10.1103/PhysRevD.110.064028",
    journal = "Phys. Rev. D",
    volume = "110",
    number = "6",
    pages = "064028",
    year = "2024"
}

@article{Begnoni:2025mtz,
    author = "Begnoni, Andrea and Del Pozzo, Walter and Pegorin, Matteo and Pomper, Joachim and Ricciardone, Angelo",
    title = "{Tests of general relativity with Einstein Telescope}",
    eprint = "2511.07520",
    archivePrefix = "arXiv",
    primaryClass = "gr-qc",
    doi = "10.1088/1475-7516/2026/09/112",
    journal = "JCAP",
    volume = "09",
    pages = "112",
    year = "2026"
}

@misc{SymDALI,
  author  = {Felipe Barbosa},
  title   = {\texttt{SymDALI}},
  version= {1.0.0},
  year   = {2026},
  howpublished  = {\url{https://github.com/Felipe-4/SymDALI}}
}

@article{Foreman-Mackey:2012any,
    author = "Foreman-Mackey, Daniel and Hogg, David W. and Lang, Dustin and Goodman, Jonathan",
    title = "{emcee: The MCMC Hammer}",
    eprint = "1202.3665",
    archivePrefix = "arXiv",
    primaryClass = "astro-ph.IM",
    doi = "10.1086/670067",
    journal = "Publ. Astron. Soc. Pac.",
    volume = "125",
    pages = "306--312",
    year = "2013"
}

@article{Vallisneri:2007ev,
    author = "Vallisneri, Michele",
    title = "{Use and abuse of the Fisher information matrix in the assessment of gravitational-wave parameter-estimation prospects}",
    eprint = "gr-qc/0703086",
    archivePrefix = "arXiv",
    reportNumber = "LIGO-P070009-00-Z",
    doi = "10.1103/PhysRevD.77.042001",
    journal = "Phys. Rev. D",
    volume = "77",
    pages = "042001",
    year = "2008"
}

@article{Rodriguez:2013mla,
    author = "Rodriguez, Carl L. and Farr, Benjamin and Farr, Will M. and Mandel, Ilya",
    title = "{Inadequacies of the Fisher Information Matrix in gravitational-wave parameter estimation}",
    eprint = "1308.1397",
    archivePrefix = "arXiv",
    primaryClass = "astro-ph.IM",
    doi = "10.1103/PhysRevD.88.084013",
    journal = "Phys. Rev. D",
    volume = "88",
    number = "8",
    pages = "084013",
    year = "2013"
}

@article{Mapelli:2021gyv,
    author = "Mapelli, Michela and Bouffanais, Yann and Santoliquido, Filippo and Sedda, Manuel Arca and Artale, M. Celeste",
    title = "{The cosmic evolution of binary black holes in young, globular, and nuclear star clusters: rates, masses, spins, and mixing fractions}",
    eprint = "2109.06222",
    archivePrefix = "arXiv",
    primaryClass = "astro-ph.HE",
    doi = "10.1093/mnras/stac422",
    journal = "Mon. Not. Roy. Astron. Soc.",
    volume = "511",
    number = "4",
    pages = "5797--5816",
    year = "2022"
}

@article{ET:2025xjr,
    author = "Abac, Adrian and others",
    collaboration = "ET",
    title = "{The Science of the Einstein Telescope}",
    eprint = "2503.12263",
    archivePrefix = "arXiv",
    primaryClass = "gr-qc",
    reportNumber = "ET-0036C-25",
    doi = "10.1088/1475-7516/2026/03/081",
    journal = "JCAP",
    volume = "03",
    pages = "081",
    year = "2026"
}

@article{Evans:2021gyd,
    author = "Evans, Matthew and others",
    title = "{A Horizon Study for Cosmic Explorer: Science, Observatories, and Community}",
    eprint = "2109.09882",
    archivePrefix = "arXiv",
    primaryClass = "astro-ph.IM",
    reportNumber = "CE-P2100003-v7",
    month = "9",
    year = "2021"
}

@article{Sellentin:2014zta,
    author = "Sellentin, Elena and Quartin, Miguel and Amendola, Luca",
    title = "{Breaking the spell of Gaussianity: forecasting with higher order Fisher matrices}",
    eprint = "1401.6892",
    archivePrefix = "arXiv",
    primaryClass = "astro-ph.CO",
    doi = "10.1093/mnras/stu689",
    journal = "Mon. Not. Roy. Astron. Soc.",
    volume = "441",
    number = "2",
    pages = "1831--1840",
    year = "2014"
}

@article{LIGOScientific:2026fcf,
    author = "Abac, A. G. and others",
    collaboration = "LIGO Scientific, VIRGO, KAGRA",
    title = "{GWTC-4.0: Tests of General Relativity. II. Parameterized Tests}",
    eprint = "2603.19020",
    archivePrefix = "arXiv",
    primaryClass = "gr-qc",
    reportNumber = "LIGO-P2500066",
    month = "3",
    year = "2026"
}

@article{Begnoni:2025oyd,
    author = "Begnoni, Andrea and Anselmi, Stefano and Pieroni, Mauro and Renzi, Alessandro and Ricciardone, Angelo",
    title = "{Detectability and Parameter Estimation for Einstein Telescope Configurations with GWJulia}",
    eprint = "2506.21530",
    archivePrefix = "arXiv",
    primaryClass = "astro-ph.CO",
    month = "6",
    year = "2025"
}

@article{LIGOScientific:2025obp,
    author = "Abac, A. G. and others",
    collaboration = "LIGO Scientific, Virgo, KAGRA",
    title = "{Black Hole Spectroscopy and Tests of General Relativity with GW250114}",
    eprint = "2509.08099",
    archivePrefix = "arXiv",
    primaryClass = "gr-qc",
    reportNumber = "LIGO P2500461",
    doi = "10.1103/6c61-fm1n",
    journal = "Phys. Rev. Lett.",
    volume = "136",
    number = "4",
    pages = "041403",
    year = "2026"
}

@article{Santoliquido:2025lot,
    author = "Santoliquido, Filippo and others",
    title = "{Fast and accurate parameter estimation of high-redshift sources with the Einstein Telescope}",
    eprint = "2504.21087",
    archivePrefix = "arXiv",
    primaryClass = "astro-ph.HE",
    doi = "10.1103/wf1k-p5cl",
    journal = "Phys. Rev. D",
    volume = "112",
    number = "10",
    pages = "103015",
    year = "2025"
}

@article{Hu:2024mvn,
    author = "Hu, Qian and Veitch, John",
    title = "{Costs of Bayesian parameter estimation in third-generation gravitational wave detectors: An assessment of current acceleration methods}",
    eprint = "2412.02651",
    archivePrefix = "arXiv",
    primaryClass = "gr-qc",
    reportNumber = "ET-0666A-24",
    doi = "10.1103/dj7k-tk37",
    journal = "Phys. Rev. D",
    volume = "112",
    number = "8",
    pages = "084039",
    year = "2025"
}

@article{Santoliquido:2025aiq,
    author = "Santoliquido, Filippo and Tissino, Jacopo and Dupletsa, Ulyana and Branchesi, Marica and Harms, Jan",
    title = "{Comparing next-generation detector configurations for high-redshift gravitational wave sources with neural posterior estimation}",
    eprint = "2512.20699",
    archivePrefix = "arXiv",
    primaryClass = "gr-qc",
    doi = "10.1051/0004-6361/202558828",
    journal = "Astron. Astrophys.",
    volume = "708",
    pages = "A175",
    year = "2026"
}

@article{Wette:2020air,
    author = "Wette, Karl",
    title = "{SWIGLAL: Python and Octave interfaces to the LALSuite gravitational-wave data analysis libraries}",
    eprint = "2012.09552",
    archivePrefix = "arXiv",
    primaryClass = "astro-ph.IM",
    doi = "10.1016/j.softx.2020.100634",
    journal = "SoftwareX",
    volume = "12",
    pages = "100634",
    year = "2020"
}

@software{nessai,
   author       = {Michael J. Williams},
   title        = {nessai: Nested Sampling with Artificial Intelligence},
   month        = feb,
   year         = 2021,
   publisher    = {Zenodo},
   version      = {latest},
   doi          = {10.5281/zenodo.4550693},
   url          = {https://doi.org/10.5281/zenodo.4550693}
 }

@article{bilby_paper,
    author = "Ashton, Gregory and others",
    title = "{BILBY: A user-friendly Bayesian inference library for gravitational-wave astronomy}",
    eprint = "1811.02042",
    archivePrefix = "arXiv",
    primaryClass = "astro-ph.IM",
    doi = "10.3847/1538-4365/ab06fc",
    journal = "Astrophys. J. Suppl.",
    volume = "241",
    number = "2",
    pages = "27",
    year = "2019"
}

@software{bilby_doi,
    author       = {Colm Talbot and
                    Gregory Ashton and
                    Moritz Hübner and
                    Matt Pitkin and
                    plasky and
                    Michael J. Williams and
                    asb5468 and
                    Aditya Vijaykumar and
                    Rory Smith and
                    SMorisaki and
                    John Veitch and
                    Nikhil Sarin and
                    Duncan Macleod and
                    Daniel Williams and
                    JasperMartins and
                    MarcArene and
                    C P L Berry and
                    Vivien Raymond and
                    Ceciliogq and
                    Ivan Markin and
                    David Keitel and
                    AlexandreGoettel and
                    Lorenzo Pompili and
                    Mick Wright and
                    oliviawilk and
                    noahewolfe and
                    jacobgolomb and
                    Shichao Wu and
                    Rhiannon Udall and
                    Michael Pürrer},
    title        = {bilby-dev/bilby: v.2.7.1},
    month        = nov,
    year         = 2025,
    publisher    = {Zenodo},
    version      = {v2.7.1},
    doi          = {10.5281/zenodo.17533961},
    url          = {https://doi.org/10.5281/zenodo.17533961},
    swhid        = {swh:1:dir:3840032346ac7f56582e6c74cc48dccc33747418
                    ;origin=https://doi.org/10.5281/zenodo.14025463;vi
                    sit=swh:1:snp:b4b0318c71303094707fd61e3305dd2addf1
                    a743;anchor=swh:1:rel:7b67a76d1e7e224899d337492665
                    5fa8e0e7074c;path=bilby-dev-bilby-710e180
                    },
    }

@article{Emma:2026urt,
    author = "Emma, Mattia and Ashton, Gregory",
    title = "{Residual neural likelihood estimation and its application to gravitational-wave astronomy}",
    eprint = "2601.13857",
    archivePrefix = "arXiv",
    primaryClass = "gr-qc",
    doi = "10.1103/f5df-cxyg",
    journal = "Phys. Rev. D",
    volume = "113",
    number = "12",
    pages = "124064",
    year = "2026"
}

\end{document}